%% file: main.tex
\documentclass[10pt, conference, compsocconf]{IEEEtran}
\ifCLASSINFOpdf
\else
\fi
\newcommand{\depth}{$\mathit{D}$}
\newcommand{\num}{$\mathit{T}$}

\newcommand{\reg}{$\mathit{R}$}
\newcommand{\bank}{$\mathit{B}$}

\newcommand{\depthM}{\mathit{D}}
\newcommand{\numM}{\mathit{T}}

\newcommand{\bankM}{\mathit{B}}

\newcommand{\name}{DPU-v2}

\usepackage{color}

\usepackage{soul}
\newcommand{\rtwo}[1]{#1}
\newcommand{\re}[1]{#1}

\newcommand{\ns}[1]{{\color{red}[ns: #1]}}
\newcommand{\mv}[1]{{\color{green} #1}}
\newcommand{\wm}[1]{{\color{cyan}[wm: #1]}}

\soulregister{\ns}{1}
\soulregister{\mv}{1}
\soulregister{\wm}{1}
\newcommand\mi[1]{\mathit{#1}}

\newcommand{\mycap}[1]{\textmd{#1}}

\usepackage[ruled,vlined,linesnumbered]{algorithm2e}

\usepackage[pagebackref=true, hidelinks]{hyperref} 
\renewcommand*\backref[1]{\ifx#1\relax \else (Cited on #1) \fi}

\usepackage[frozencache,cachedir=.]{minted}
\AtBeginEnvironment{minted}{%
  \catcode`?\active
  \begingroup\lccode`~=`\?\lowercase{\endgroup\def~{\linebreak}}%
}

\usepackage{amssymb}

\usepackage{multirow}
\usepackage{cite}
\usepackage{booktabs}
\usepackage{graphicx}

\let\labelindent\relax
\usepackage[shortlabels]{enumitem}

\begin{document}
%
\title{\name: Energy-efficient execution of irregular directed acyclic graphs}

\author{\IEEEauthorblockN{Nimish Shah\IEEEauthorrefmark{1},
Wannes Meert\IEEEauthorrefmark{2} and
Marian Verhelst\IEEEauthorrefmark{1}}
\IEEEauthorblockA{\IEEEauthorrefmark{1}Department of Electrical Engineering - MICAS, KU Leuven, Belgium
}
\IEEEauthorblockA{\IEEEauthorrefmark{2}
Department of Computer Science - DTAI, KU Leuven, Belgium\\
Emails: nshah@esat.kuleuven.be, wannes.meert@kuleuven.be, marian.verhelst@kuleuven.be}
}



\maketitle
\thispagestyle{plain}
\pagestyle{plain}

\begin{abstract}
A growing number of applications like probabilistic machine learning, sparse linear algebra, robotic navigation, etc., exhibit \textit{irregular} data flow computation that can be modeled with directed acyclic graphs (DAGs). The irregularity arises from the seemingly random connections of nodes, which makes the DAG structure unsuitable for vectorization on CPU or GPU. Moreover, the nodes usually represent a small number of arithmetic operations that cannot amortize the overhead of launching tasks/kernels for each node, further posing challenges for parallel execution.

To enable energy-efficient execution, this work proposes DAG processing unit (DPU) version 2, a \textbf{specialized processor architecture} optimized for irregular DAGs \rtwo{with static connectivity}. It consists of a tree-structured datapath for efficient data reuse, a customized banked register file, and interconnects tuned to support irregular register accesses. \name{} is utilized effectively through a \textbf{targeted compiler} that systematically maps operations to the datapath, minimizes register bank conflicts, and avoids pipeline hazards. Finally, a \textbf{design space exploration} identifies the optimal architecture configuration that minimizes the energy-delay product. This hardware-software co-optimization approach results in a speedup of 1.4$\times$, 3.5$\times$, and 14$\times$ over a state-of-the-art DAG processor ASIP, a CPU, and a GPU, respectively, while also achieving a lower energy-delay product. In this way, this work takes an important step towards enabling an embedded execution of emerging DAG workloads.
\end{abstract}


%
\IEEEpeerreviewmaketitle

\section{Introduction} \label{sec:introduction}
\begin{figure}[!t]
\centering
\includegraphics[trim={0cm 0cm 0cm 0cm} , clip, width=0.98\columnwidth]{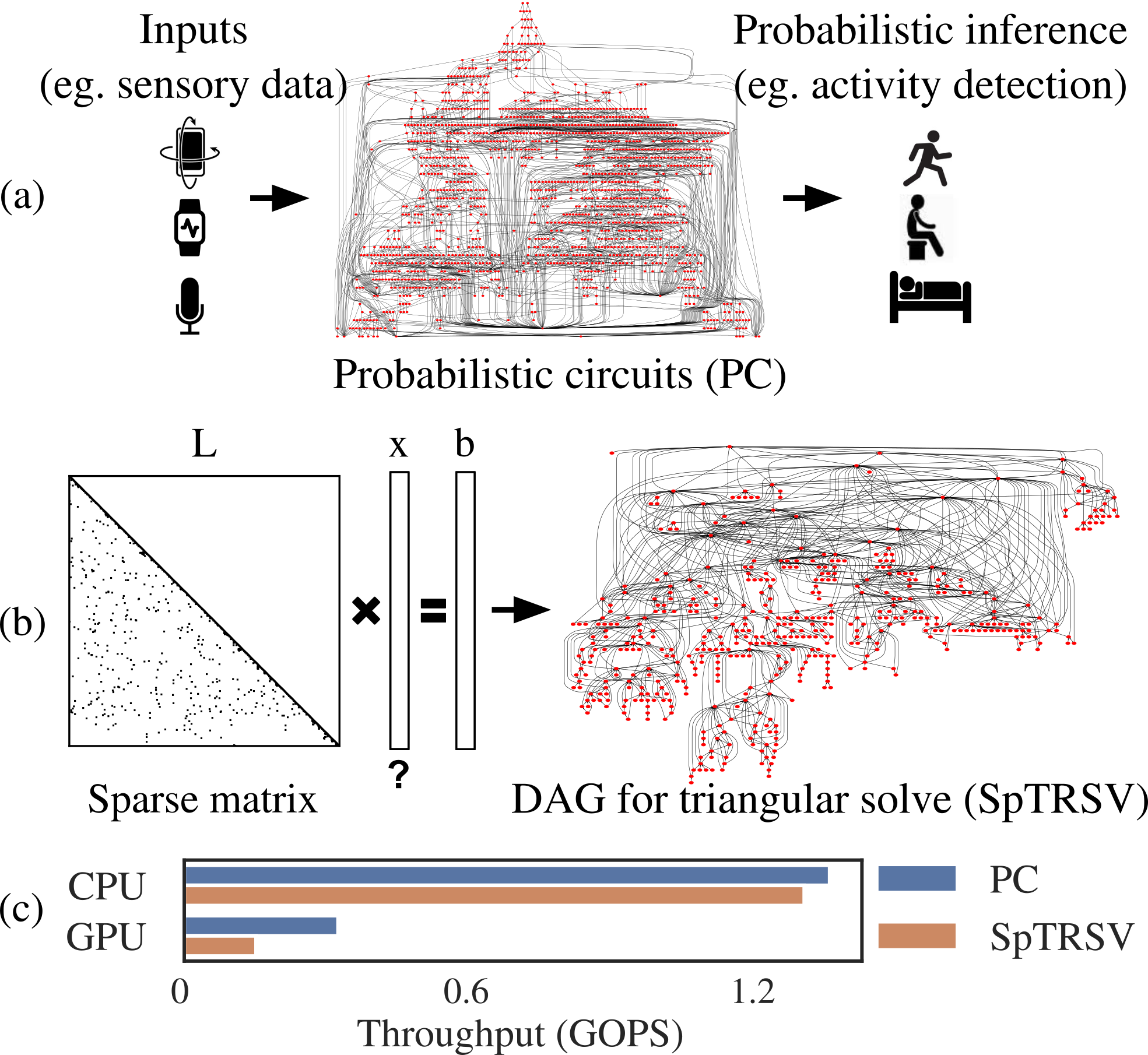}
\caption{\mycap{Example applications with irregular computation DAGs. (a) Probabilistic circuits (PC) used in machine learning, and (b) Sparse matrix triangular solves (SpTRSV) used in linear algebra. (c) CPU and GPU achieve significantly lower than their peak throughput for such DAGs.}}%
\label{fig:sparse_applications}
\vspace{-5pt}
\end{figure}

Emerging machine learning models like probabilistic circuits (PC, also called sum-product networks) \cite{choi2020probabilistic,poon2011sum} are increasingly used for energy-constrained applications like robotic navigation \cite{8967568}, robust image classification \cite{DBLP:conf/icml/StelznerPK19}, human activity recognition \cite{galindez2019towards}, etc. PCs exhibit an irregular flow of data that can be represented with directed acyclic graphs (DAG), as shown in fig.~\ref{fig:sparse_applications}(a) for a human activity recognition model. Similarly, sparse linear algebra operations like sparse triangular solve (SpTRSV) used for embedded applications like robotic localization and mapping \cite{polok2016accelerated}, wireless communication \cite{wang2016overview}, cryptography \cite{delaplace2018linear}, etc., can also be represented with irregular DAGs. Fig.~\ref{fig:sparse_applications}(b) shows an example of a DAG for a matrix from the SuiteSparse collection \cite{DBLP:journals/toms/DavisH11}. In these DAGs, nodes represent arithmetic operations and edges represent the dependencies among these operations.



Unfortunately, these irregular \re{unstructured} DAGs pose challenges to general-purpose computing architectures like GPUs and single-instruction-multiple-data (SIMD) vectorization in CPUs.
 Fig. \ref{fig:sparse_applications}(c) shows the throughputs of an Intel Xeon CPU and an Nvidia RTX GPU for these applications.
 The CPU performs much below the peak achievable throughput (3.4 TOPS in this case), and the GPU even underperforms compared to the CPU despite having highly parallel hardware (especially for DAGs with fewer than 100K nodes). This underperformance is caused by the following two reasons, preventing the platforms from utilizing the sparsity in the workloads:
\begin{itemize}[itemsep=0em, leftmargin=*]
    \item The seemingly-random \re{edge connectivity of an unstructured DAG} results in irregular memory accesses with poor spatial locality. Hence, the large granularity of cache-line size leads to severe under-utilization of caches and memory bandwidth, because only 4B might get consumed out of the 32/64B of data fetched to a cache line. Furthermore, these irregular accesses also prevent memory coalescing, which is crucial for high GPU performance \cite{van2015improving}.
    \item Due to the irregular structure, parallelizing different parts of DAGs across multiple units (like CPU cores, GPU streaming multiprocessors, etc.) demands high communication and synchronization overhead, limiting the parallelization benefits \cite{shah2021graphopt}.
    
\end{itemize}
 



\looseness=-1 To address these issues, this paper proposes 
DAG processing unit (\name{}), a processing architecture\footnote{\re{Available at https://github.com/nimish15shah/DAG\_Processor.}} optimised for irregular DAGs along with a custom-designed targeted compiler. 
The major contributions of this work are as follows:

\begin{enumerate}[itemsep=0em, leftmargin=*]
    \item \textbf{Processor:} A parameterized architecture template is designed with a tree-structured datapath for immediate reuse of data, a banked register file with a novel addressing scheme, and high-bandwidth flexible interconnects. The architecture is made programmable with a custom instruction set with long words.
    
    \item \textbf{Compiler:} 
    A specialized compiler is designed that schedules DAG nodes such that the tree-based datapath is maximally utilized. The irregularity of DAGs leads to frequent register bank conflicts, which are reduced with a conflict-aware register allocation while taking into account the interconnect constraints.
    
    \item \textbf{Design-space exploration:} Since there are multiple design parameters in the architecture template (e.g. the number of register banks, computational parallelism, etc.), a design space exploration is performed to arrive at the architecture with the minimum energy-delay product in a 28nm CMOS technology, and compared with state-of-the-art 
    implementations. 
    
\end{enumerate}

\rtwo{This work focuses on DAGs that remain static across multiple executions, which allow amortization of the compilation overhead, enabling highly targeted (but relatively slow) DAG-specific compilation. This assumption of DAGs being static is valid for our target workloads: (1) the structure of a trained PC remains static during inference, and (2) in many applications using SpTRSV like robotic localization and mapping \cite{polok2016accelerated}, wireless communication \cite{wang2016overview}, cryptography \cite{delaplace2018linear}, etc., the sparsity pattern of a matrix remains static while the numerical values or the right-hand side vector change across executions, which effectively only changes the inputs of the DAG.
Furthermore, DAGs are assumed to contain only arithmetic nodes (and no conditional nodes) that are executed in a predefined order based on edge-induced dependencies. This assumption is also valid for the target workloads in this paper, but excludes operations like breadth-first search (even on a static DAG) from the scope of this work.}
    
The paper is organized as follows. \S \ref{sec:challenges} discusses this work's approach for efficient parallel execution of DAGs. \S \ref{sec:hardware} describes the processor architecture template of \name{} followed by \S \ref{sec:compiler} explaining the compiler. Next, \S \ref{sec:experiments} presents the design-space exploration to find the design with the minimum energy-delay product, and compares it with state-of-the-art work. Finally, \S \ref{sec:related_works} discusses related works and \S \ref{sec:conclusion} concludes the paper.

\section{Concepts to tame irregularity}
\label{sec:challenges}
In this paper, a DAG encodes the computations to be performed for an application and its data dependencies. A DAG node represents arithmetic operations (e.g. additions, multiplications) to be performed on the node's inputs. \re{This work focuses on DAGs with \textit{fine-grained} nodes with low arithmetic intensity. In other words, the nodes represent a scalar or a small-sized vector operation but not large tensor operations, in contrast to DAGs of deep neural networks, for example, which have compute-intensive tensor operations as nodes.}
The output result of a node serves as an input to \re{(possibly multiple)} nodes connected to the outgoing edges. 

\begin{figure}[!t]
\centering
\includegraphics[trim={0cm 0cm 0cm 0cm} , clip, width=0.98\columnwidth]{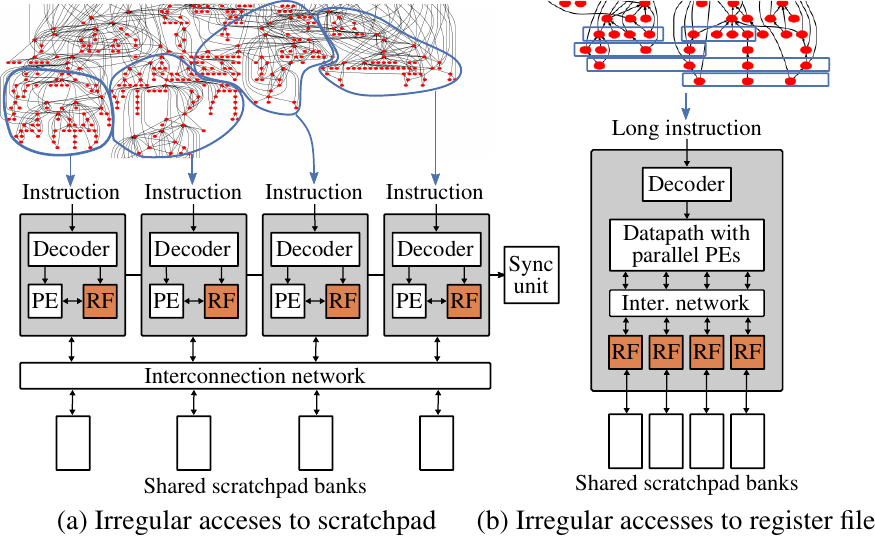}
\caption{\mycap{(a) A common approach for executing DAGs in parallel, in which the unpredictable irregular accesses happen to scratchpad/memory. (b) In \name{} by pushing the interconnect closer to the datapath, the compiler is equipped to predict the irregular accesses and prevent bank conflicts. 
}}%
\label{fig:async_vs_sync}
\vspace{-10pt}
\end{figure}%
\vspace{5pt}
\noindent \textbf{Execution order}: Executing a DAG means evaluating all the nodes in a \textit{valid} order imposed by the directed edges representing data dependencies. For all the edges, the source node should be executed before the destination node. Put differently, a node can be executed only after all the \textit{predecessor} nodes 
have finished execution. \re{Note that the DAGs targeted in this work do not contain conditional branch nodes, and hence all the nodes are supposed to be executed.}

\vspace{5pt}
\noindent \textbf{What can be executed in parallel?} 
At a given moment, all the nodes whose predecessor nodes have finished execution can be executed in parallel. The execution begins with the source nodes of the DAG, as they do not have any incoming edges. Subsequently, different nodes become ready as the execution progresses. DAGs from real-world applications typically have ample parallelism, which can be utilized through a parallel processing architecture.

The rest of this section discusses the challenges for parallel processing of irregular DAGs and the techniques introduced in this work to address them.


\subsection{Making irregular data accesses predictable} \label{sec:async_vs_sync}
Fig.~\ref{fig:async_vs_sync}(a) shows a common approach for accelerating DAGs, in which a DAG is partitioned into smaller subgraphs that can be executed on parallel cores, e.g, in a multicore CPU. This approach is used in works like \cite{picciau2016balancing}, \cite{shah2021graphopt}, \cite{shah2021dpu}. The cores have to synchronize occasionally for race-free communication that happens through a shared memory structure like an L3 cache in CPUs or a shared scratchpad in \cite{shah2021dpu}. Due to the DAG irregularity, the accesses to this shared memory structure usually happen to random addresses, inducing frequent bank conflicts in the typically-used banked memory structures. These conflicts could become a severe bottleneck. For instance, in \cite{shah2021dpu}, 43\% of the load requests result in bank conflicts. Work in \cite{dadu2019towards}, \cite{rucker2021capstan} alleviates the impact of these conflicts through aggressive reordering of requests, but incurs the overhead of a reordering buffer. Yet, none of these works attempt to avoid the conflicts altogether. Given that the DAG structure is known at compile time, it could be possible to perform the memory allocation such that the simultaneous requests do not access the same bank. The main challenge for such 
conflict avoidance is that it is not straightforward to predict which requests happen simultaneously. The parallel cores normally execute \textit{asynchronously}, i.e. a stall in one core (eg., to wait for the next instruction) does not stall the others, preventing a compiler from predicting conflicts.

Fig.~\ref{fig:async_vs_sync}(b) shows an alternative approach that makes the simultaneous irregular accesses predictable by pushing the interconnect closer to the datapath and forcing the PE to operate in a \textit{synchronous} way. This limits the irregular accesses 
to the register file, while the memory/scratchpad can be accessed in a regular pattern.
In every cycle, the compiler knows which registers are accessed and can thereby also predict register bank conflicts. The datapath with parallel PEs, however, cannot be programmed with SIMD-like instructions because the PEs will gather data from different addresses in the register banks owing to DAG irregularity. Instead, VLIW (very long instruction word)-like instructions can be used. Yet, unlike conventional VLIW with a partitioned register file \cite{capitanio1992partitioned}, the datapath does not have to be limited to an array of PEs, but complex dataflow patterns can be utilized. 

This work hypothesizes that the architecture in fig.~\ref{fig:async_vs_sync}(b) can improve throughput by avoiding bank conflicts and by using a datapath tuned for the target workloads.

\subsection{Reducing data accesses} \label{sec:systolic_vs_tree}
\begin{figure}[!t]
\centering
\includegraphics[trim={0cm 0cm 0cm 0cm} , clip, width=0.8\columnwidth]{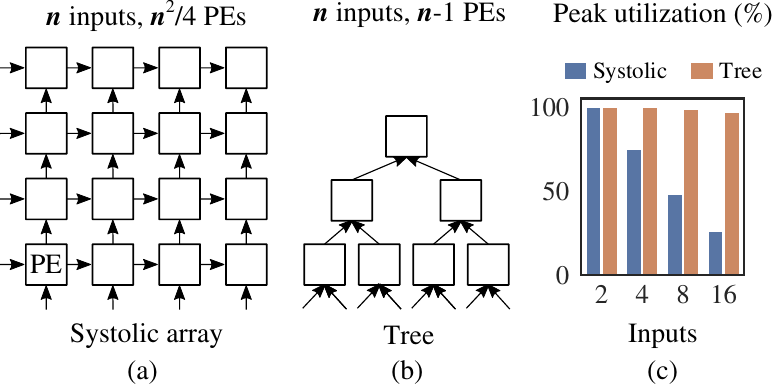}
\caption{\mycap{Systolic arrays are under-utilized by irregular DAGs, while a tree-shaped datapath is a promising alternative.
}
}%
\label{fig:systolic_vs_tree}
\vspace{-10pt}
\end{figure}

To ease the problem of bank conflicts further and to reduce energy consumption,
the intermediate results in the computation could be immediately consumed in the datapath, avoiding reads/writes from a register file.
A commonly used topology for such data reuse is a systolic array of PEs \cite{kung1979systolic} (fig.~\ref{fig:systolic_vs_tree}(a)), yet it is unsuitable for sparse DAGs due to severe hardware under-utilization as described below. 


\vspace{5pt}
\noindent\textbf{Peak utilization}: 
To check the suitability of the systolic array, the spatial datapath mapper from \cite{nowatzki2014scheduling} is used to find the largest subgraph in our target DAGs (described in \S \ref{sec:workloads}) that can be mapped to the array. This essentially is the peak utilization possible for the datapath. Fig.~\ref{fig:systolic_vs_tree}(c) shows that the peak utilization of the systolic array drops rapidly with the increasing number of inputs, and even a 4$\times$4 array with 8 inputs cannot be properly utilized.


\vspace{5pt}
\noindent\textbf{Tree of PEs}: The tree-structured datapath (fig.~\ref{fig:systolic_vs_tree}(b)) is a promising alternative candidate, as multi-input DAG nodes can be spatially unrolled as binary trees. 
As shown in fig.~\ref{fig:systolic_vs_tree}(c), the spatial mapper confirms that there exist subgraphs in DAGs that can fully utilize the datapath.

The spatial mapper used here can quantify the peak utilization, but the constrained-optimization-based approach used in the mapper is too slow to fully map large DAGs with tens of thousands of nodes. For a scalable solution, this work presents a heuristic algorithm to decompose irregular DAGs into tree-like subgraphs (\S \ref{sec:block_decomposition}).



\subsection{Supporting irregular data accesses}\label{sec:motivate_crossbar}
\begin{figure}[!t]
\centering
\includegraphics[trim={0cm 0cm 0cm 0cm} , clip, width=0.7\columnwidth]{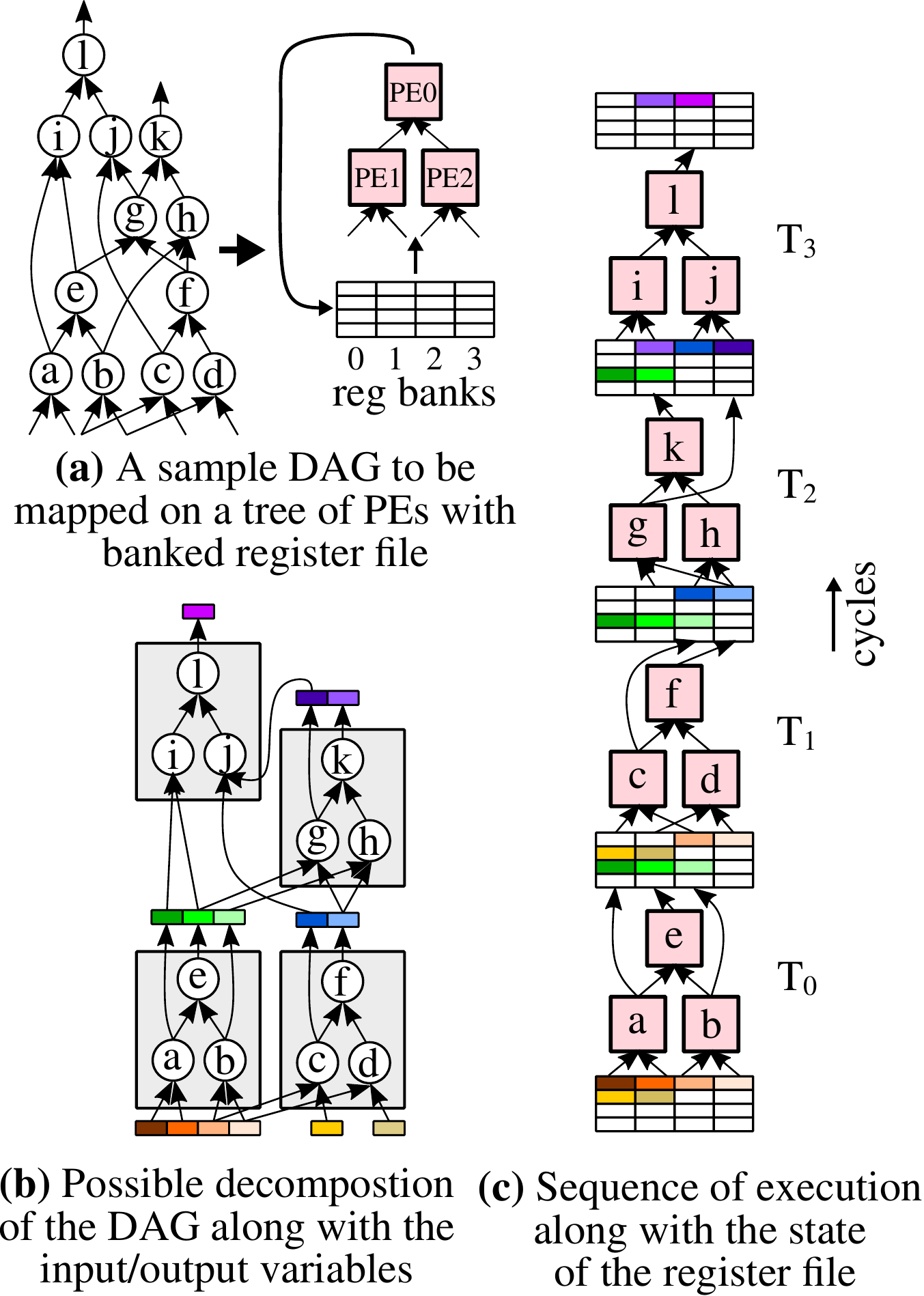}
\caption{\mycap{Example of a DAG execution on a tree of PEs, causing irregular register accesses. 
}}%
\label{fig:motivate_crossbar}
\vspace{-10pt}
\end{figure}

Section \S \ref{sec:async_vs_sync} alluded that DAG irregularity causes irregular register accesses. Fig.~\ref{fig:motivate_crossbar} illustrates this with a concrete example.
Fig.~\ref{fig:motivate_crossbar}(a) shows a sample DAG and a target datapath consisting of a PE tree and a banked register file for inputs/outputs. Fig.~\ref{fig:motivate_crossbar}(b) shows an example decomposition of the DAG in tree-shaped structures that can be sequentially mapped to the datapath. The outputs of some of the nodes (like node $a$) are required to be stored in the register file for later consumption, while others (like node $d$) are fully consumed within the datapath.
Fig.~\ref{fig:motivate_crossbar}(c) shows the temporal execution with the evolving state of the register file. The DAG irregularity manifests in the form of irregular accesses to register banks as explained next.

\begin{figure*}[!t]
\centering
\includegraphics[trim={0cm 0cm 0cm 0cm} , clip, width=0.9\textwidth]{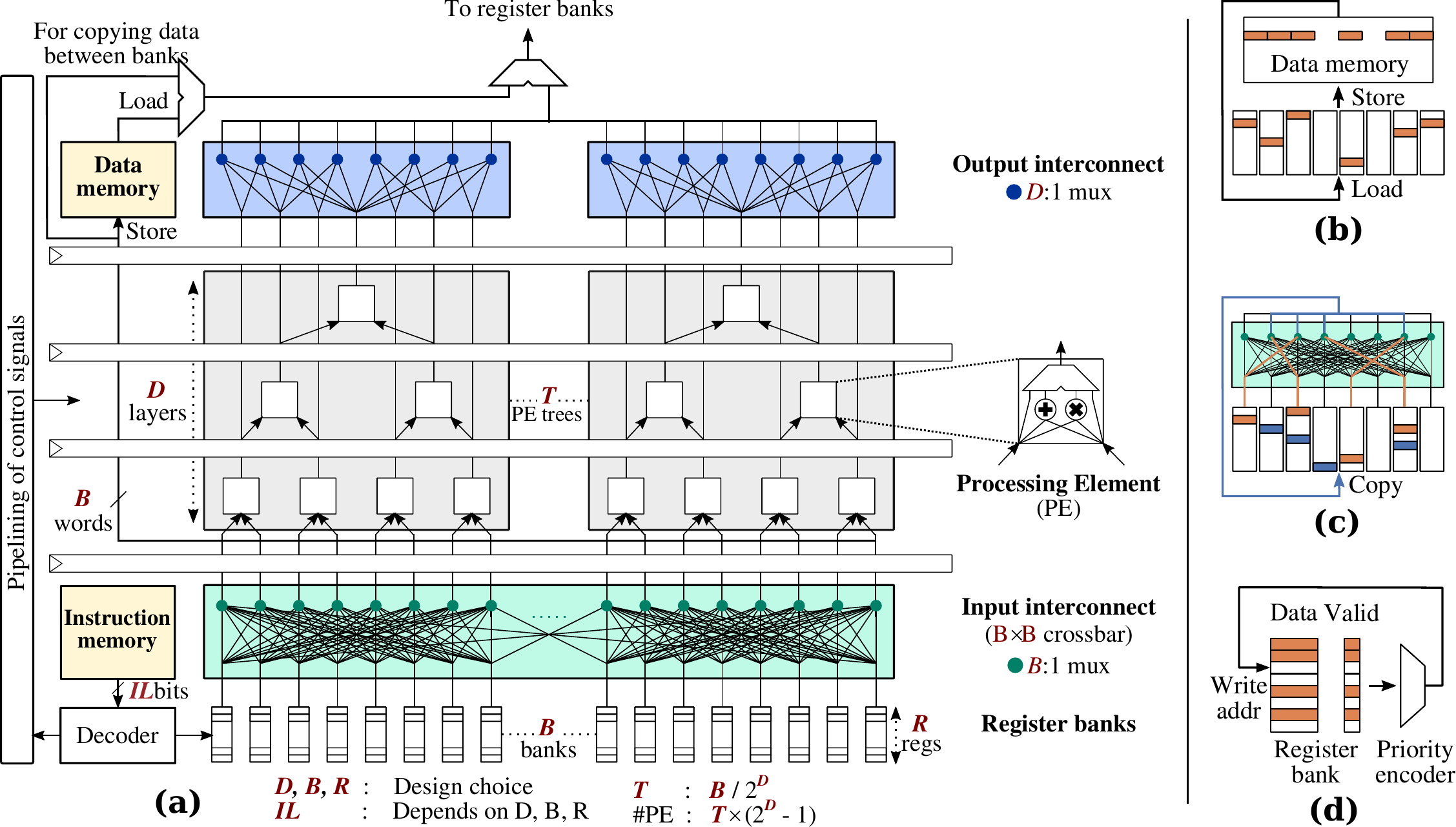}
\caption{\mycap{(a) \textbf{\name{} architecture template} consists of processing elements (PE) connected in a tree topology. The trees are connected to parallel register banks with input and output interconnects. The datapath is pipelined according to the number of PE layers. 
(b) Vectors are loaded/stored from a single address in the data memory, but register banks are addressed independently. 
(c) To avoid conflicts, the data are copied between banks via the input interconnect.
(d) Write address generation by tracking the occupancy status with valid bits. 
}}%
\label{fig:processor_arch}
\vspace{-10pt}
\end{figure*}%

\vspace{5pt}
\noindent \textbf{Irregular bank accesses}: The inputs and outputs of the PE tree access different banks over the cycles. For example, PE2 reads banks 2 and 3 at T$_0$ and banks 1 and 3 at T$_1$. The access pattern cannot be \textit{regularized} by simply remapping the variables to different banks or remapping nodes to different PEs. Consider the input variables of node $b$, which are also consumed by $c$ and $d$. At T$_0$, they are consumed by one PE, while at T$_1$, by two PEs, one each. It is not possible to map the variables among banks such that the PEs read the same banks at T$_0$ and T$_1$ both. One alternative could be to execute the nodes $c$ and $d$ in different cycles, but that would lead to idle PEs. Hence, for high PE utilization, some form of flexibility is required in interconnecting the inputs/outputs of PE trees to the register banks.

\vspace{5pt}
\noindent \textbf{Irregular addresses}: PEs read different addresses from different banks in a cycle. 
This is caused by the fact that the input variables consumed simultaneously can possibly be generated in different cycles. (eg., the inputs of T$_3$). 
Similarly, variables that are generated together can be consumed in different cycles (eg., the outputs of T$_1$). Thus, unlike SIMD execution, all the register banks do not use the same read/write addresses, requiring flexible addressing hardware.

\vspace{5pt}
\noindent \textbf{Replication is not a solution}: It may appear that replicating a variable in multiple banks and multiple addresses can solve both of the above problems for register reads. 
For such replications, a variable would be written as well as read multiple times. However, there are three problems:
\begin{itemize}[itemsep=0em, leftmargin=*]
  \item The register file can be severely underutilized due to replicated copies of the same data. 
  \item The energy consumption increases due to more writes compared to fig.~\ref{fig:motivate_crossbar}'s approach of single writes.
  \item The replication alleviates the irregular reads but exacerbates the irregularity of writes. Hence, it simply shifts the problem from reads to writes.
\end{itemize}

Hence, this work aims to use flexible interconnects and independent register addresses to ease the gathering of data from irregular locations without relying on data replication, enabling high-throughput execution. 

\section{\name{} Architecture template} \label{sec:hardware}

This section describes the proposed processor architecture template 
(fig.~\ref{fig:processor_arch}(a)) designed for unstructured DAGs. 



\subsection{Parallel tree of PEs}
As shown in fig. \ref{fig:processor_arch}(a), the datapath consists of many parallel processing elements (PEs) to execute the DAG nodes.

\begin{itemize} [itemsep=0em, leftmargin=*]
    \item \textbf{PE}: Each PE 
    can be configured to perform a basic arithmetic operation ($+$ and $\times$). Additionally, a PE can bypass one of its inputs to its output.
    \item \textbf{Trees of PEs}: PEs are interconnected in \num{} parallel structures with a tree topology containing \depth{} layers, where outputs of a PE layer are connected to the inputs of the next PE layer. This enables immediate reuse of intermediate results, avoiding register file accesses and the associated energy consumption.
    \item \textbf{Pipeline stages}: The PE outputs are registered, resulting in \depth{} pipeline stages in the trees.
 
\end{itemize}





\subsection{Shared register file}
The PE trees read and write data from a shared register file containing \bank{} parallel banks with \reg{} registers per bank (fig. \ref{fig:processor_arch}(a)). The number of banks \bank{} 
is chosen such that there is one register bank for each input of the trees (i.e., $\bankM{} = \numM{}\times\mathrm{2}^{\depthM{}}$). As such, 
the register file has enough bandwidth to feed the data into the trees in every cycle.
However, due to DAG irregularity, the parallel inputs needed in a clock cycle typically do not reside at the same address in the different banks, as discussed in \S \ref{sec:motivate_crossbar}. To handle this irregularity, the banks are made \textit{independent} in terms of read/write addressing. Each bank can read/write from/to any location, independent of the other banks. This flexibility comes at the overhead of additional instruction bits to encode these addresses. While this overhead is present for reads, it is 
alleviated
for writes as discussed below.


\vspace{5pt}
\noindent \textbf{Automatic write address generation}: \label{sec:automatic_writes}
To reduce the instruction size, a novel register-writing policy is used, alleviating the need of encoding write addresses in instructions. The policy is to always write data to the \textit{empty} location with the lowest address within a register bank. Thus instructions do not control where a write happens, but the register bank itself chooses the appropriate location for the incoming data. 

\vspace{5pt}
\noindent \textbf{Why does the policy work?} To perform correct execution with such an automatic write policy, the compiler should be able to predict at compile time the write addresses that will be chosen at run time. 
Since our target DAGs are known at compile time and the PEs execute synchronously, the 
instruction execution sequence is deterministic. This enables the compiler to predict write addresses that will be chosen in every cycle, given that the execution begins with a known state of the register banks (e.g, all the banks are empty).



\vspace{5pt}
\noindent \textbf{Hardware}: The policy requires that the occupancy state of every register is tracked. This is done via a \textit{valid} bit for every register, which is set to 1 when the corresponding register contains valid data. Using these valid bits, the write address is computed with a priority encoder as shown in fig.~\ref{fig:processor_arch}(d). 
A write to a register sets the respective valid bit. However, a read should not automatically reset the bit, as a variable residing in a register can be reused multiple times. The instructions should therefore indicate the last read to the register 
to reset the valid bit and letting the register be subsequently available to write new data. This resetting is done by a \textit{valid\_rst} bit in the instruction, one for each bank.
Thus, a multi-bit write address is replaced with a single \textit{valid\_rst} bit, reducing the width of the instructions. This automatic write policy results in a program size reduction of 30\% on average.


\subsection{Datapath-register banks connections}
\label{sec:interconnect_dse}
\begin{figure}[!t]
\centering
\includegraphics[trim={0cm 0cm 0cm 0cm} , clip, width=0.98\columnwidth]{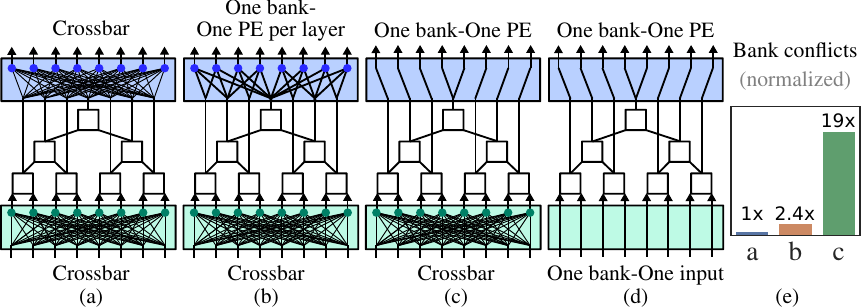}
\caption{\mycap{Different interconnection topologies and their impact on bank conflicts.
}}%
\label{fig:interconnect_dse}%
\vspace{-10pt}
\end{figure}

As discussed in \S \ref{sec:motivate_crossbar}, the connection of datapath 
to the register banks requires some flexibility due to the irregularity of DAG edges.
However, the design space for interconnecting \bank{} read ports to \bank{} inputs and \#PE outputs to \bank{} write ports is huge. On one extreme, full crossbars can be used for both inputs and outputs (fig.~\ref{fig:interconnect_dse}(a)). On the other extreme, one-to-one connections can be done (fig.~\ref{fig:interconnect_dse}(d)).

\vspace{5pt}
\noindent\textbf{Compilation considerations}: The interconnections not only have a hardware impact but also impacts the compilation complexity. Crossbars significantly simplify the compilation as it decouples PE mapping and bank mapping. For example, for the design in fig.~\ref{fig:interconnect_dse}(a), when a DAG node is mapped to a PE, the bank mapping options for its input data and output results do not get restricted, as all the banks are accessible for reading and writing. In contrast, for fig.~\ref{fig:interconnect_dse}(d), when a DAG node is mapped to a PE, its output result can be stored to only one bank (or two in the case of the top PE), and the input data should be mapped to the banks the PE can access. Thus, the PE mapping and bank mapping get coupled. This coupling complicates the compilation because the mapping of one node to a PE can impact the mapping of all the rest of the nodes.

\looseness=-1 Thus, using at least one crossbar, either at the input or the output, ensures that the mapping of one node only impacts a limited set of nodes. Furthermore, hardware synthesis results revealed that a crossbar only consumes 4\% area and 9\% power of the whole design (see table~\ref{tab:area_power_breakdown} (input interconnect)). As a result, designs with at least one crossbar are considered for exploration (fig.~\ref{fig:interconnect_dse}(a)-(c)) simplifying compilation without significant hardware overhead. The mapping algorithm is explained in detail in \S \ref{sec:pe_bank_allocation}, which is used to minimize the bank conflicts (fig.~\ref{fig:interconnect_dse}(e)) for the different design options. The bank conflicts happen due to the following structural hazards: (1) two PEs simultaneously access the same bank (but banks only have one read and one write port), or (2) the input data (or the output result) is mapped to a bank that the PE cannot access, inducing a stall for copying data across banks. 

As expected, the design (a) achieves minimal conflicts. Yet, the design (b) is selected, which has a crossbar for inputs, and each bank is connected to outputs of one PE per layer. It induces only 1.4$\times$ higher conflicts due to the limited output connectivity, increasing the overall latency by 1\% but reducing the overall power by 9\%. As such, it achieves a better latency-power trade-off. 
The design (c) results in significantly higher conflicts without considerable area/power advantage over (b). 
The design (d) is not evaluated further as it will incur even more conflicts than (c).

\subsection{Load, store, and copy of data}
The register banks connect to an on-chip SRAM-based data memory with a read-write width of \bank{} words. The load/store from the data memory happens in the form of vectors (with a word-level enable mask) as shown in fig.~\ref{fig:processor_arch}(b).
But, this data can be written to/read from different addresses in the register banks.
Note that the register write addresses for a load operation are automatically generated as described in \S \ref{sec:automatic_writes}, whereas the register read addresses for a store operation are encoded in the instructions. Furthermore, to handle bank conflicts, a copy instruction enables an arbitrary shuffle of data across banks using the crossbar as shown in fig.~\ref{fig:processor_arch}(c),
implemented as depicted on the left of fig.~\ref{fig:processor_arch}(a).


\subsection{Long, variable length instructions}
\begin{figure}[!t]
\centering
\includegraphics[trim={0cm 0cm 0cm 0cm} , clip, width=0.98\columnwidth]{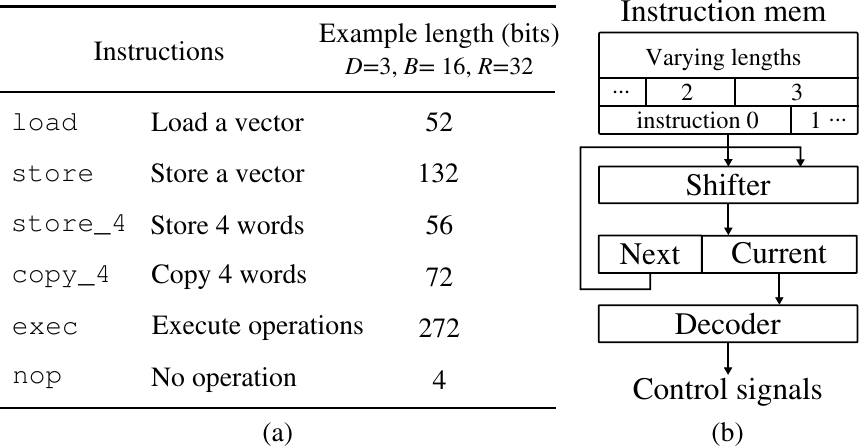}
\caption{\mycap{Variable-length instructions (in (a)) are densely packed in the memory (in (b)) and are appropriately aligned during decoding for stall-free execution}}%
\label{fig:decoder}%
\vspace{-10pt}
\end{figure}

The architecture is designed to be programmable for arbitrary DAGs, with a custom VLIW instruction set (fig.~\ref{fig:decoder}(a)) that can configure the trees and crossbar, copy data from one register bank to another, or load/store from data memory. As different instructions encode different types of information, they have different lengths, depending on the hardware parameters $D$, $B$, $R$. Fig.~\ref{fig:decoder}(a) shows instruction lengths for a sample design configuration.
For proper utilization of the instruction memory, the instructions are packed densely, without any bubbles (fig.~\ref{fig:decoder}(b)). The instruction memory can supply \emph{IL} bits in every cycle, which is the same as the size of the longest instruction.
To handle varying lengths, a shifter before the instruction decoder performs appropriate alignment, which makes sure there are no stalls in execution, and the next instruction is always fully available for decoding.

\looseness=-1 In summary, \name{} contains a datapath with parallel PE trees connected to a banked register file with flexible interconnects. It is programmed with custom variable-length VLIW instructions, which are packed densely in the instruction memory.
The independent parameters of the architecture template (\depth{}, \bank{} and \reg{}) are not fixed at this point and will be selected based on a design space exploration described in \S \ref{sec:experiments}.


\begin{figure*}
\centering
\includegraphics[trim={0cm 0cm 0cm 0cm} , clip, width=0.9\textwidth]{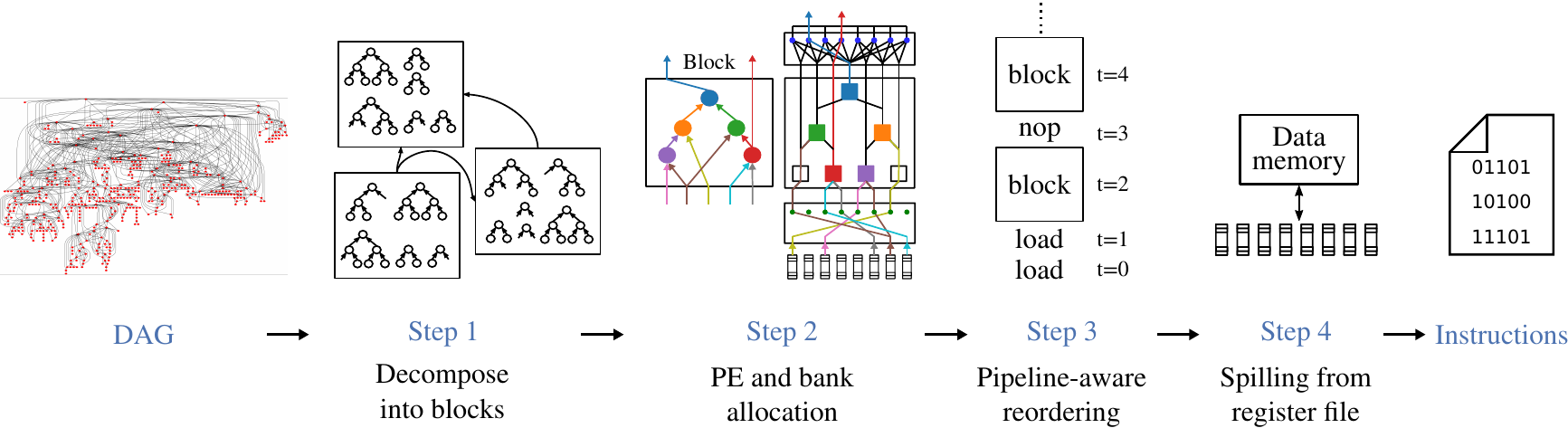}
\caption{\mycap{Steps involved in the custom compiler to generate optimized instructions for a given DAG.}}
\label{fig:compilation}
\vspace{-10pt}
\end{figure*}

\begin{figure}[!t]
\centering
\includegraphics[trim={0cm 0cm 0cm 0cm} , clip, width=0.98\columnwidth]{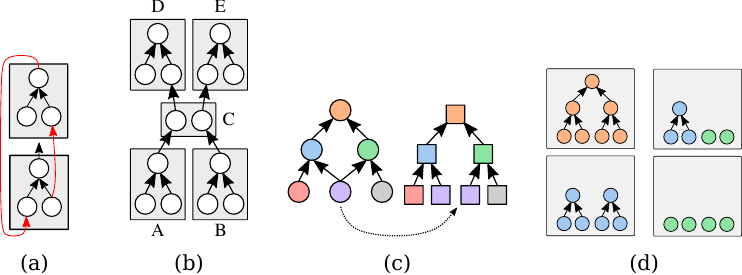}
\vspace{-5pt}
\caption{\mycap{(a) Cycles to be avoided during block generation. (b) Avoid block dependencies like the dependence of D on B. \rtwo{(c) Any subgraph with 2-input nodes, one sink node, and with the longest path length less than the depth of a PE tree can be mapped to the PE tree by replicating certain nodes. (d) Subgraphs that can be combined into a block (possibly with permutations) for mapping to a datapath with one PE tree of depth 3.}}}%
\label{fig:compiler_step_1}%
\vspace{-10pt}
\end{figure}%

\section{Compiler for DAG} \label{sec:compiler}

\looseness=-1 Due to the non-conventional datapath with flexible interconnects and the VLIW instruction set, it is difficult to modify a conventional compiler of traditional processors to generate instructions for the proposed architecture. Hence, a targeted compiler is developed to fully utilize the PE trees despite the irregularity of unstructured DAGs. The compiler is designed to work for any values of the design parameters (i.e. \depth{}, \bank{} and \reg{}), instead of targeting a fixed configuration of the architecture template. The compiler takes as input a DAG in any of the popular graph formats (i.e. all formats supported by the NetworkX package \mbox{\cite{hagberg2008exploring}}) and generates an execution binary that can be directly programmed to \name{}. \re{Since the structure of the DAG remains static across multiple executions (during inference with PC and for SpTRSV with multiple right-hand side vectors), the compilation is performed offline and the DAG is unfolded at compile time to generate DAG-specific instructions.}
The major compilation steps are shown in fig. \ref{fig:compilation} and discussed next.

\subsection{Block decomposition (Step 1)} \label{sec:block_decomposition}
Compilation begins by decomposing the input DAG, which is first converted to a binary DAG (containing 2-input nodes only) by replacing a multi-input node with a tree of 2-input nodes. This is to ensure that the nodes can be mapped directly to the 2-input PEs. This binary DAG is decomposed into sets of 
nodes called \textit{`blocks'}. A block is supposed to be a monolithic unit that can be executed with a single \texttt{exec} instruction (from fig.~\ref{fig:decoder}(a)). The constraints and objectives for this decomposition step are as follows:
\begin{itemize}[itemsep=0em, leftmargin=*]
    \item \textit{Constraint A}: The resulting graph of blocks should be acyclic for functional correctness. Fig.~\ref{fig:compiler_step_1}(a) shows an example of a cyclic decomposition, in which it cannot be determined which block should be scheduled first.
    \item \textit{Constraint B}: The nodes in a block should be spatially schedulable on the PE trees, considering the number of PEs available and their connectivity. 
    \item \textit{Objective C}: The PE trees should be maximally utilized.
    \item \textit{Objective D}: The number of 
    dependencies among blocks should be minimized to reduce read-after-write hazards during the pipelined execution. Fig.~\ref{fig:compiler_step_1}(b) shows an example where the dependency of block D on B could be avoided by an improved combination of nodes in C, as the nodes in D are not actually dependent on B in the DAG.
\end{itemize}
To achieve these objectives under the mentioned constraints, a greedy iterative algorithm (see algo.~\ref{alg:step_1}) is designed that constructs one block in every iteration as follows,
\begin{enumerate}[itemsep=0em, leftmargin=*]
    \item \textbf{Search all the \textit{schedulable} connected subgraphs.} 
    A subgraph is schedulable if all of its predecessor nodes (if any) are already assigned to the blocks in previous iterations. Ensuring this schedulability property in every iteration results in blocks that satisfy constraint~A. \re{Furthermore, a subgraph is schedulable only if it can be completely mapped to a PE tree, which is needed for constraint~B.} 
    \rtwo{Subgraphs to be considered for this schedulability check are constructed as follows: A node that is less than \depth{} distance away from the already mapped nodes (called $\mi{curr\_source\_nodes}$) is considered as a sink node, and along with its unmapped ancestors, form a subgraph. The check that whether a subgraph satisfies constraint~B or not is simplified due to the tree topology of PEs. Any connected subgraph with only 2-input nodes, exactly one sink node, and with the longest path length less than \depth{} can be mapped to a PE tree of depth \depth{}. Fig.~\ref{fig:compiler_step_1}(c) shows an example of a non-tree subgraph that satisfies these properties, and can be mapped through replication. In this way, the find\_schedulable\_subg() function (line 6) finds all the schedulable subgraphs that have a given predecessor node as an input.
    
    The overall search for subgraphs is made efficient by keeping track of the schedulable subgraphs across iterations (in the set $\mi{\mathcal{D}_{sch}}$),
    to avoid re-searching the same subgraphs again. In every iteration, only the subgraphs that originate from the nodes mapped in the previous iteration are added to the set $\mi{\mathcal{D}_{sch}}$ (lines 5-7), by setting $\mi{curr\_source\_nodes}$ appropriately (line 23).}
    
    \item \textbf{Among the schedulable subgraphs, select a set to form a block.} \rtwo{The previous step finds connected subgraphs that can be mapped to the datapath. As shown in fig.~\ref{fig:compiler_step_1}(d), multiple such subgraphs (with possibly different longest path lengths) can be combined into a block. To find the appropriate set of subgraphs to combine in a block, the \textit{fitness} of a block is evaluated based on objectives~C and D: (1) According to obj.~C, the blocks with more nodes are more fit. (2) Quantifying obj~D is difficult. In practice, the dependencies across blocks decrease if the subgraphs combined in a block lie closer to each other in the DAG.} Hence, a block is penalized according to the \textit{distance} between the nodes in a block. The distance is approximated by the difference in occurrences of their nodes during a depth-first traversal of the DAG (performed once at the beginning). \rtwo{These two metrics of fitness are used to add an appropriate subgraph to a block (lines 15-16) and to compare different valid blocks (lines 17-19).}
    
\end{enumerate}

The process repeats until all the nodes are mapped to blocks. Note that nodes are not mapped to specific PEs yet, which is done subsequently in step 2.

\rtwo{\textbf{Asymptotic complexity}: The number of iterations scale linearly with the number of DAG nodes for a given datapath. In each iteration, selecting a set of subgraphs to form a block scales linearly with the number of schedulable subgraphs (i.e. the size of $\mi{\mathcal{D}_{sch}}$), which in turn scales linearly with the number of DAG nodes in the worst case. As such, the complexity is $O$($N^2$), where $N$ is the number of nodes.}
\input{algo_step1}

\subsection{PE and register bank mapping (Step 2)} \label{sec:pe_bank_allocation}
To generate an instruction for each block, every node of the block needs to be assigned to a hardware PE in the PE trees, and source/destination of the inputs/outputs of a block should be mapped to register banks. 
The constraints and objectives of the mappings are as follows:
\begin{itemize}[itemsep=0em, leftmargin=*]
    \item \textit{Constraint E}: Only one node can be mapped to one PE (note that the replication in \ref{fig:compiler_step_1}(c) is achieved by splitting nodes into multiple temporary nodes). The mapping should be topologically consistent, which requires that if a node $n$ is mapped to a PE $p$, then the predecessors and successors of $n$ are also mapped to, respectively, the predecessors and successors of $p$. 
    \item \textit{Constraint F}: Two different inputs of a block should not be mapped to the same register bank, to prevent bank conflicts while reading.
    \item \textit{Constraint G}: Two different outputs of a block should not be mapped to the same register bank, to prevent bank conflicts while writing.
    \item \textit{Constraint H}: For the nodes whose results are stored as outputs of a block, the PE mapping and the bank mapping of that node should be compatible, i.e. the bank should be writable from that PE. This constraint arises due to the limited connectivity of PEs and banks in the output interconnect (fig. \ref{fig:processor_arch}(a)). Note that such a constraint is not required for block inputs because the input interconnect is a crossbar.
    \item \textit{Objective I}: Minimize bank conflicts, as every bank conflict will result in a stalling cycle for data copy.
    \item \textit{Objective J}: Balance the distribution across register banks. 
\end{itemize}

\input{algo_step2}

\begin{figure}[!t]
\centering
\includegraphics[trim={0cm 0cm 0cm 0cm} , clip, width=0.8\columnwidth]{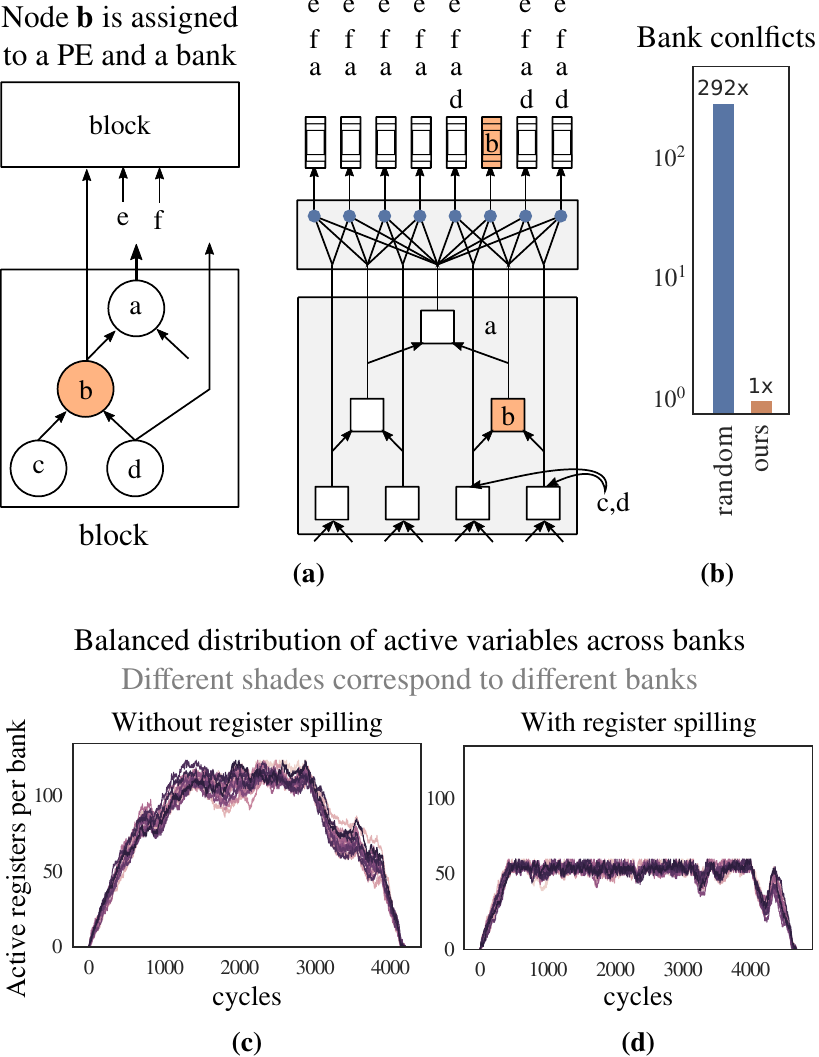}
\caption{\mycap{(a) An example of how the mapping of node $b$ affects the compatible PEs and banks of the other nodes. 
(b) Our algorithm achieves considerably lower bank conflicts than random allocation.
(c) and (d) shows that the data allocation across register banks remains well balanced.
}}%
\vspace{0pt}
\label{fig:bank_allocation}%
\end{figure}

Due to constraint H, the PE mapping is performed in tandem with the register bank mapping, since assigning a node to PE restricts the number of compatible banks and vice versa. Specifically, a greedy iterative algorithm is used which performs the mappings for one node in every iteration \rtwo{as described in algorithm~\ref{alg:step_2}}. The algorithm keeps track of a set of \textit{compatible} PEs ($S_p$) for every node and a set of \textit{compatible} banks ($S_b$) for nodes that serve as inputs/outputs of blocks (and hence should be stored to the register file). \rtwo{These nodes involved in inputs/outputs (e.g., nodes $a$, $b$, and $d$ in fig.~\ref{fig:bank_allocation}(a)) are called $\mi{io\_nodes}$ in the algorithm}. \re{A PE is compatible if the topological consistency defined in constraint E can be satisfied for all the nodes in the block after the node is mapped to the PE. A bank is compatible, if mapping the node to that bank does not cause a bank conflict with the already mapped nodes.} 

\noindent\textbf{Node order for mapping}:
\rtwo{The mapping is done for $io\_nodes$ first}. In every iteration, the node with the least number of compatible banks (to achieve objective $I$) in $S_b$ is chosen for mapping, from anywhere in the DAG ignoring the block boundaries. \rtwo{The runtime of the selection is made independent of the size of the DAG using the $\mi{\mathcal{M}_{nodes}}$ datastructure (constructed in lines 9-12 in algo. \ref{alg:step_2}), and searching in it in every iteration (lines 15-18).}

\noindent\textbf{Mapping}:
\rtwo{The chosen node is mapped to a compatible bank (chosen randomly to achieve objective J) if available, otherwise to a bank that leads to the least conflicts (lines 21-24). Subsequently, the node is mapped to a PE (lines 25-29) that can write to the chosen bank, if such a PE is compatible. If not, a compatible PE is chosen at random from $S_p$.} Note that at least one compatible PE will always be left because, as explained in step 1(\S \ref{sec:block_decomposition}), the blocks satisfy constraint~B, ensuring the schedulability of nodes. 
Fig. \ref{fig:bank_allocation}(a) shows the mapping of a node $b$ to an appropriate PE and bank (in color). The following two updates happen to the $S_p$ and $S_b$ of the nodes affected by this mapping.

\noindent \textbf{Intra-block compatibility update}: 
The PE and bank options for the unmapped nodes $a$, $c$ and $d$ decrease due to the mapping of $b$ in fig.~\ref{fig:bank_allocation}(a), according to the constraints~E and~G. $S_p$ and $S_b$ for these nodes are updated accordingly \rtwo{(lines 30-33)}.

\noindent \textbf{Inter-block compatibility update}: Nodes from other blocks are also affected by the bank assignment to $b$. The nodes $e$ and $f$ in fig.~\ref{fig:bank_allocation}(a) are consumed together with $b$ and cannot be allocated to the same bank (to satisfy constraint~F). Hence, $b$'s bank is removed from their $S_b$ \rtwo{(line 34)}. 

After these compatibility updates, a new node is selected in the next iteration, and the process repeats until all the $\mi{io\_nodes}$ are mapped. \rtwo{Next, the non $\mi{io\_nodes}$ (node $c$ in fig.~\ref{fig:bank_allocation}(a)) are mapped to PEs in the same way as discussed earlier, but without the steps for bank mapping. 

\textbf{Asymptotic complexity}: The work done in the intra-block update depends only on the size of the datapath and does not depend on the properties of the DAG, while the work done in the inter-block update scales with the outdegree of the node being assigned. Since the number of iterations is same as the number of nodes, the algorithm complexity is $O$($N \times \Delta (G)$), where $N$ is the number of DAG nodes and $\Delta (G)$ is the maximum outdegree of the DAG.}

Fig.~\ref{fig:bank_allocation}(b) shows that our algorithm achieves a 292$\times$ reduction in bank conflicts compared to random bank allocation, confirming that the objective~$I$ is met. Fig.~\ref{fig:bank_allocation}(c) shows the number of active registers across banks during the execution of a workload (from table~\ref{tab:workload_details}), demonstrating a well-balanced distribution, in-line with objective~$J$. The spilling of registers for limited bank size is discussed later in this section. The bank conflicts are resolved with the \texttt{copy} instruction from fig.~\ref{fig:decoder} that copies the data to an appropriate bank, eliminating the need for handling the bank conflicts in hardware.
Finally, an instruction list is constructed with the \texttt{exec} and \texttt{copy} instructions, along with the \texttt{load} instructions to fetch inputs from the data memory for the first time. 

\textbf{Impact of the crossbar}:
Note that the input crossbar significantly simplifies the inter-block update. It decouples the PE allocations of a block from the bank allocations of its inputs, and limits constraint H to outputs only. This ensures that the inter-block updates remain limited to the inputs of the successor blocks only, and do not propagate to the whole DAG. Without a crossbar, the inter-block update can possibly affect all the nodes in the DAG, quickly reducing the number of compatible banks for the rest of the nodes in every iteration.



\subsection{Pipeline-aware reordering (Step 3)} \label{sec:pipelining}
As the datapath has $D+1$ pipestages, the instruction list (generated in step 2) is re-ordered to ensure that the dependent instructions are at least $D+1$ instructions apart to prevent read-after-write (RAW) hazards. The reordering involves a search for independent instructions to insert in between dependent instructions. This search is limited to a fixed-size window of succeeding instructions (300 in our experiments), to make the runtime scale linearly with the size of the instruction list. Subsequently, no-operation (\texttt{nop}) instructions are inserted for the unresolved hazards.


\renewcommand{\arraystretch}{0.9}
\begin{table}[]
\centering
\caption{\mycap{\re{\rtwo{Statistics of the benchmarked DAGs}}}} \label{tab:workload_details}
\resizebox{0.48\textwidth}{!}{%
\begin{tabular}{lclcccc}
\toprule
& \multicolumn{1}{c}{Type} & \multicolumn{1}{c}{Workloads} & \begin{tabular}[c]{@{}c@{}}Nodes \\ (n) \end{tabular} & \begin{tabular}[c]{@{}c@{}}Longest \\ path (l)\end{tabular} & \begin{tabular}[c]{@{}c@{}} n/l \end{tabular} &
\begin{tabular}[c]{@{}c@{}}\rtwo{Compile} \\ \rtwo{time (min.)}\end{tabular}
\\ \midrule

\multirow{6}{*}{\textbf{(a)}}

& \multirow{6}{*}{PC}
 & tretail    & 9k   & 49 & 180 & 1\\
 & & mnist      & 10k  & 26 & 400 & 2\\
 & & nltcs      & 14k  & 27 & 504 & 3\\
 & & msnbc      & 48k  & 28 & 1.7k & 28\\
 & & msweb      & 51k  & 73 & 698 & 31\\
 & & bnetflix   & 55k  & 53 & 1k & 35\\

 \midrule
 \multirow{6}{*}{\textbf{(b)}}
& \multirow{6}{*}{SpTRSV} 
& bp\_200   & 8k   & 139  & 60 & 1\\
& & west2021 & 10k  & 136  & 74 & 1\\
& & sieber   & 23k  & 242  & 94 & 6\\
& & jagmesh4 & 44k  & 215  & 203 & 12\\
& & rdb968   & 51k  & 278  & 181 & 15\\
& & dw2048   & 79k  & 929  & 85 & 16\\

 \midrule
\multirow{4}{*}{\textbf{(c)}}
& \multirow{4}{*}{\re{Large PC}}
 &  \re{pigs}    & 0.6M   & 90 & 6.9k & 121\\
& & \re{andes}    & 0.7M   & 84 & 8.6k & 145\\
& & \re{munin}    & 3.1M   & 337 & 9.1k & 850\\
& & \re{mildew}    & 3.3M   & 176 & 18.8k & 910\\

\bottomrule
\end{tabular}%
}%
\end{table}

\begin{figure*}[!t]
\centering
\includegraphics[trim={0cm 0cm 0cm 0cm} , clip, width=1.65\columnwidth]{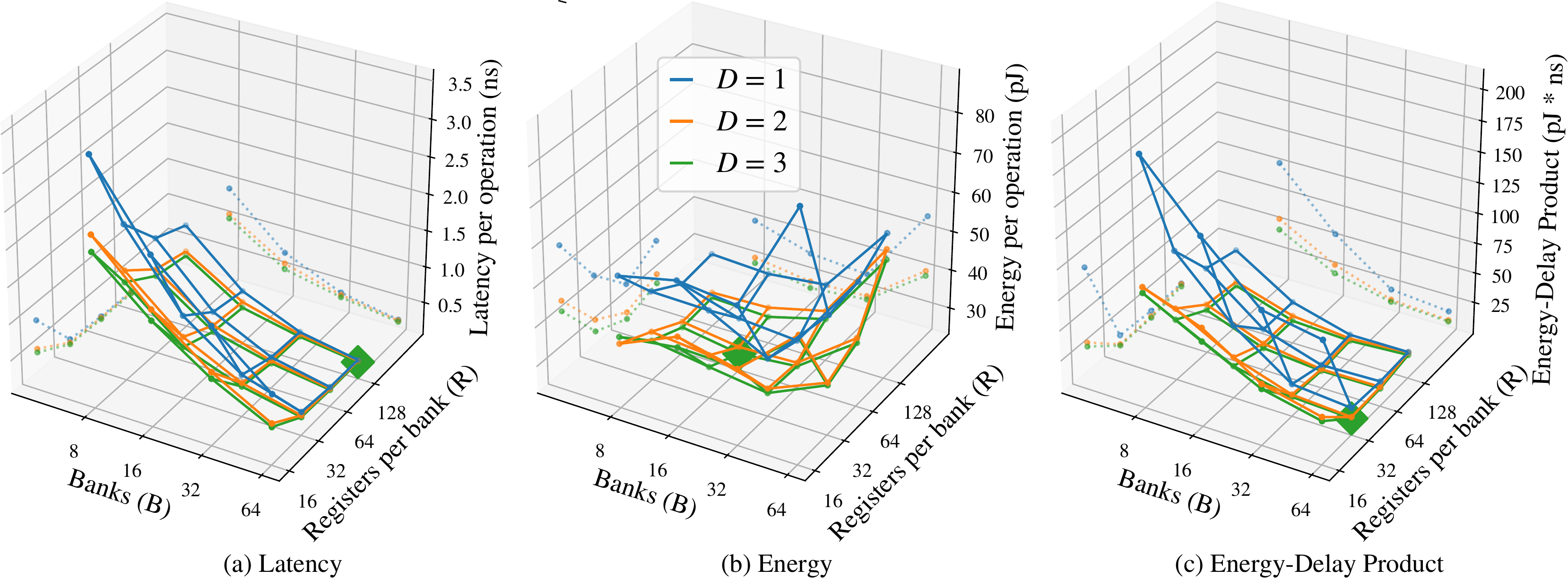}
\caption{\mycap{Design space exploration to identify the min-EDP design. The optimal points are highlighted with large markers.
}}%
\vspace{-5pt}
\label{fig:dse}%
\end{figure*}
\subsection{Spilling from register file (Step 4)}
Data must be spilled from the register file if intermediate results do not fit. Given the schedule of execution, a live-range analysis is performed to determine when the spilling is required, and \texttt{store} instructions are appropriately inserted. 
The spilled data are later loaded back by inserting \texttt{load} instructions before they are supposed to be consumed, in a way that
avoid new RAW pipeline hazards. \rtwo{The live-range analysis and insertion of intermediate loads/stores scale linearly with the size of the DAG.}
Fig.~\ref{fig:bank_allocation}(c) and (d) shows register occupancy profile for a workload without and with register spilling, respectively, for \reg{}=64. 

The final output of the compilation process is a list of instructions to be executed on \name{} for the given DAG, which can be directly translated into a binary program for execution. The distribution of different types of instructions is shown in fig.~\ref{fig:instr_stat_all_w} in the next section. 
\re{
\subsection{Reduction in memory footprint} \label{sec:mem_footprint}
The list of instructions generated by the compiler statically encodes the entire DAG structure, leading to a larger instruction footprint compared to a conventional approach of a \texttt{for} loop iterating over the DAG stored in a compressed sparse row (CSR)-like format and performing indirect memory accesses. However, note that the overall memory footprint (instructions + data) of our approach is in fact lower than the CSR datastructure because the statically generated instructions can encode the edge connectivity information with fewer bits, due to the following reasons.
\begin{itemize} [itemsep=0em, leftmargin=*]
    \item For the DAG edges that are mapped to PE-PE connections, the addresses for the source and destination of the edges need not be stored, in contrast to the CSR datastructure which stores the pointers for every edge.
    \item Due to the static mapping of variables to the register file, the address for the source/destination of the edges that are mapped to PE-register or register-PE connections can be specified with a register address (= 11b in our final design configuration) instead of encoding it in a global address space of 32/64b like in the CSR datastructure.
\end{itemize}
For the target workloads, the total memory footprint of instructions and data is 48\% smaller than the CSR datastructure, which is not required in our approach.}


\section{Design space exploration} \label{sec:experiments}
Experiments are designed to find the most energy-efficient hardware configuration (i.e. \depth{}, \bank{} and \reg{}) of \name{}
for a suite of target workloads, and compare it with state-of-the-art implementations.

\subsection{Workloads}\label{sec:workloads}
The performance is benchmarked with compute DAGs of varying sizes (see table~\ref{tab:workload_details}) from two classes of workloads.

\vspace{5pt}
\noindent \textbf{Probabilistic circuits (PC).} Probabilistic circuits (also called sum-product networks) are used in machine learning for robust inference under uncertainty, safety-critical tasks, and neuro-symbolic reasoning 
    \cite{DBLP:conf/icml/StelznerPK19,8967568,galindez2019towards, thoma2021recowns}.
    A PC is an irregular DAG in which the nodes are either sum or product operations.
    The experiments are performed on standard benchmarks of density-estimation applications from \cite{DBLP:conf/uai/LiangBB17} and from \cite{shen2016tractable}.

\vspace{5pt}
\noindent \textbf{Sparse matrix triangular solves (SpTRSV).} Solving a matrix triangular system is a fundamental operation in linear algebra \cite{davis2016survey}, used in embedded applications like robotic navigation \cite{polok2016accelerated}, wireless communication \cite{wang2016overview}, cryptography \cite{delaplace2018linear}, etc. Real-world matrices usually turn out to be highly sparse and the resulting compute DAG highly irregular. Benchmarking is done on matrices of varying sizes from the SuiteSparse benchmark of real applications \cite{DBLP:journals/toms/DavisH11}.

\begin{figure}[!t]
\centering
\includegraphics[trim={0cm 0cm 0cm 0cm} , clip, width=0.9\columnwidth]{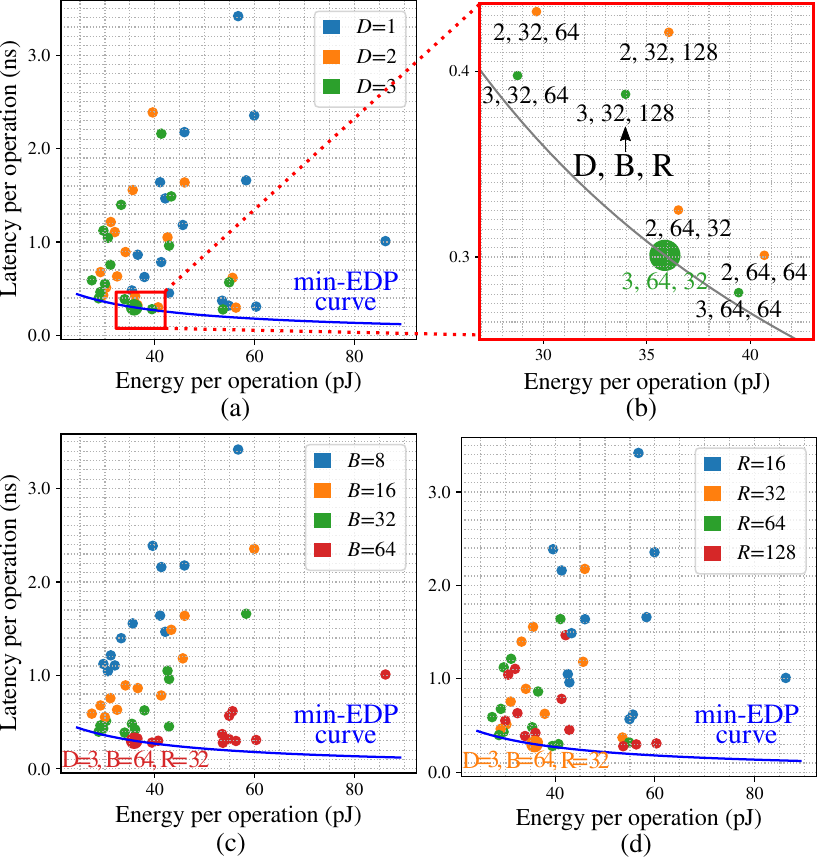}
\caption{\mycap{Latency vs. Energy plot with a constant EDP curve passing through the min-EDP point, colored according to $D$, $B$, and $R$ in (a), (c), and (d).
} \vspace{-13pt}}%
\label{fig:pareto}%
\end{figure}

\subsection{The most-efficient design configuration}\label{sec:efficient_design}

To reliably estimate the energy impact of design choices, a parameterized Verilog model is developed, which is used to synthesize gate-level netlists in a CMOS 28nm technology. The design space exploration is done by varying the parameters in the following set of values: \depth{} in [1, 2, 3], \bank{} in [8, 16, 32, 64], and \reg{} in [16, 32, 64, 128], leading to 48 combinations. For each of the design points, the energy is computed by mapping actual workloads ((a) and (b) in table \ref{tab:workload_details}) with the compiler, and annotating the resulting switching activities of the workloads from gate-level netlist simulations with a target frequency of 300MHz. The mean latency, energy, and energy-delay product (EDP) per operation, averaged over the workloads, are plotted in fig.~\ref{fig:dse} (a), (b), and (c), respectively.

\begin{table}[!t]
\centering
\caption{Area and power breakdown of \name{}}
\label{tab:area_power_breakdown}
\resizebox{0.37\textwidth}{!}{%
\begin{tabular}{lrrrrr}
\toprule
\multirow{2}{*}{} & \multicolumn{2}{c}{\textbf{Area}} & & \multicolumn{2}{c}{\textbf{Power}}  \\
 & mm$^2$ & \% & & mW & \% \\ \midrule
Datapath: &  &  &  &  &  \\
\hspace{3mm}PEs & 0.13 & 4 &  & 11.9 & 11 \\
\hspace{3mm}Pipelining registers & 0.04 & 1 &  & 8.0 & 7 \\
\hspace{3mm}Input interconnect &  0.14  & 4 &  & 10.0 & 9 \\
\hspace{3mm}Output interconnect & 0.01  & 0 &  & 0.5 & 1 \\

Register file: &  &  &  &  &  \\
\hspace{3mm}Banks & 0.35 & 11 &  & 24.0 & 22 \\
\hspace{3mm}Wr addr generator & 0.03 & 1 &  & 7.8 & 7 \\

Control: &  &  &  &  &  \\
\hspace{3mm}Instr fetch & 0.06 & 2 &  & 7.0 & 6 \\
\hspace{3mm}Decode & 0.04 & 1 &  & 2.6 & 2 \\
\hspace{3mm}Pipelining registers & 0.01 & ~0 &  & 2.7 & 2 \\

Instruction memory  & 1.20 & 38 &  & 27.7 & 26 \\
Data memory & 1.20 & 38  &  & 6.7 & 6 \\

\midrule
 & 3.20 &  & & 108.9 &  \\ \bottomrule
\end{tabular}%
}
\end{table}

As expected, the minimum latency design point (\depth{}=3, \bank{}=64, \reg{}=128) has the highest values of \depth{}, \bank{}, and \reg{}, .i.e., the one with the largest area. Note that increasing \reg{} beyond 32 has diminishing improvement as the active set of data is able to fit within the register file.
Unlike latency, the min. energy point (\depth{}=3, \bank{}=16, \reg{}=64) uses fewer \bank{} (and hence also fewer PE trees) and fewer \reg{}.
Notice that increasing \depth{} improves latency without consuming additional power, resulting in an energy improvement as well. This demonstrates the effectiveness of the tree-based spatial datapath. On the other hand, increasing \bank{} decreases latency but with proportionally more power consumption because the utilization of the increasingly parallel datapath gradually decreases, shifting the minimum energy point towards a lower value of \bank{}. 

\vspace{5pt}
\noindent \textbf{Minimum EDP:} The minimum energy-delay product (EDP, fig.~\ref{fig:dse}(c)) is achieved at (\depth{}=3, \bank{}=64, \reg{}=32). Fig.~\ref{fig:pareto} shows latency vs. energy charts for a different perspective of the design space, along with a constant-EDP curve passing through the min-EDP point. 
The slope of the curve indicates that the latency has more variation than the energy. 

\begin{figure}[!t]
\centering
\includegraphics[trim={0cm 0cm 0cm 0cm} , clip, width=0.85\columnwidth]{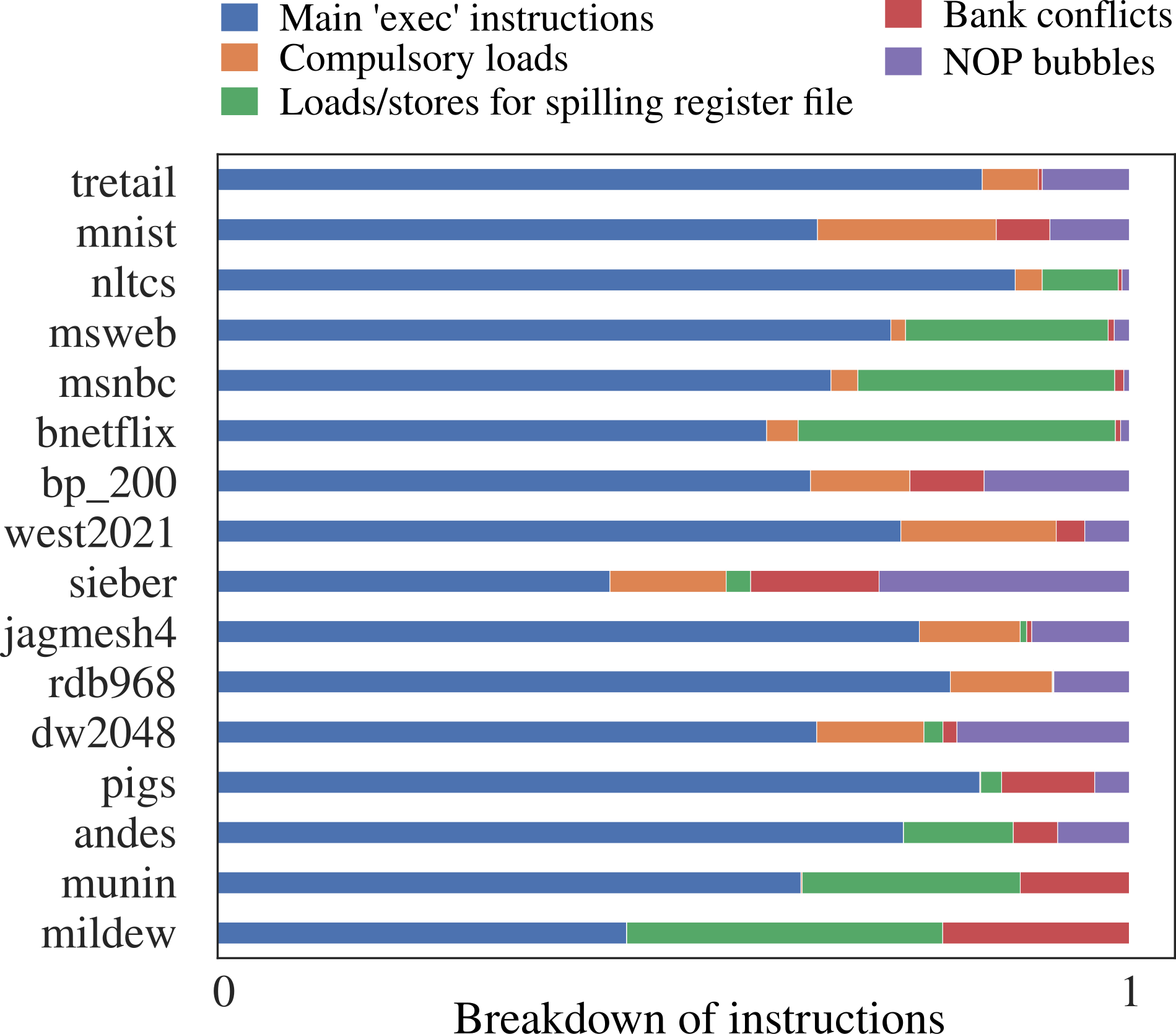}
\caption{\mycap{Breakdown of instructions.}}.%
\vspace{-10pt}
\label{fig:instr_stat_all_w}
\end{figure}

\vspace{5pt}
\noindent \textbf{Breakdown of area, power, and latency.} Table~\ref{tab:area_power_breakdown} shows the distribution of area and power across different modules for the min-EDP design. Memories consume most of the area and 32\% of the power. The datapath and the register file consume around equal power: 28\% and 29\%, respectively. Fig.~\ref{fig:instr_stat_all_w} shows the breakdown of the different categories of instructions across the different workloads executed on this architecture instantiation.

\vspace{5pt}
\rtwo{
\noindent \textbf{Compilation time}
The compilation times of the DAGs are reported in table~\ref{tab:workload_details} for the min-EDP design. For the large PCs, the decomposition into blocks (step 1) becomes too slow and the DAG is first coarsely decomposed into partitions with 20k nodes each, using the technique described in \cite{shah2021graphopt} (which scales linearly with DAG size), and then each partition is decomposed independently into blocks. 
}

\begin{figure}[!t]
\centering
\includegraphics[trim={0cm 0cm 0cm 0cm} , clip, width=0.98\columnwidth]{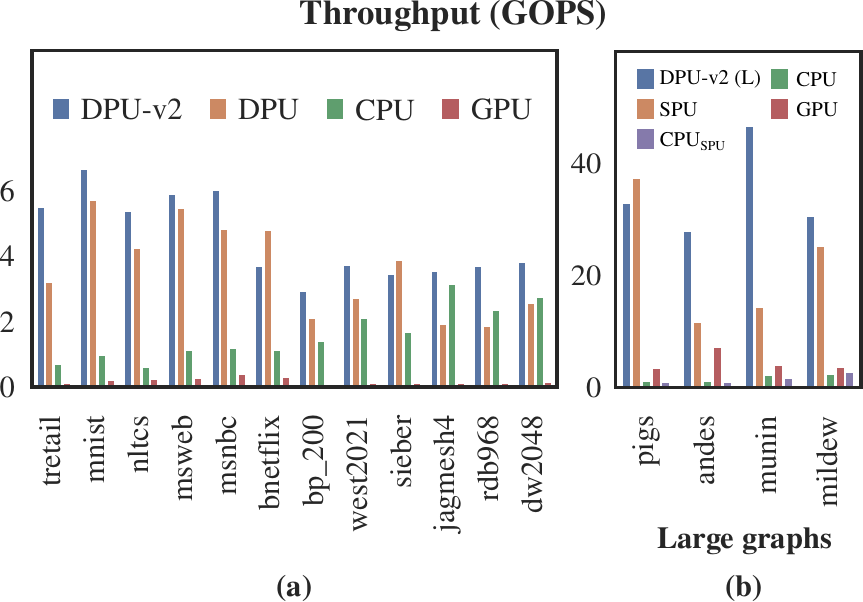}
\caption{\mycap{\re{Throughput for every workload. 
}} \vspace{-20pt}}.%
\label{fig:throughput_all_w}
\end{figure}

\begin{table*}[!t]
\centering
\caption{\mycap{\re{Performance comparison with other platforms. $^\dagger$Estimated based on the speedups over CPU$_{SPU}$.}}}%
\label{tab:sota_comparison}
\resizebox{0.9\textwidth}{!}{%
\begin{tabular}{|l|cccc|ccccc|}
\toprule
  & \textbf{\name{}}  & DPU\cite{shah2021dpu}  & CPU\cite{shah2021graphopt} & GPU \cite{naumov2011parallel} & \textbf{\name{} (L)} & SPU\cite{dadu2019towards} & CPU$_{SPU}$ \cite{dadu2019towards} & CPU\cite{shah2021graphopt} & GPU  \\\midrule
Workloads  & \multicolumn{4}{c|}{PC (a) and SpTRSV (b) from Table \ref{tab:workload_details}} & \multicolumn{5}{c|}{Large PC (c) from Table \ref{tab:workload_details}} \\ 
Technology         & \textbf{28nm       } & 28nm        & 14nm      & 12nm        & \textbf{28nm        } & 28nm           & 14nm      & 14nm        & 12nm  \\
Area (mm$^2$)      & \textbf{3.2        } & 3.6         & NA        & 754         & \textbf{40.4        } & 36.6           & NA        & NA          & 754  \\
Freq (GHz)         & \textbf{0.3        } & 0.3         & 3         & 1.35        & \textbf{0.3         } & NA             & 3         & 3           & 1.35  \\

Mem BW (GB/s)      & \textbf{--         } & --          & 120       & 616         & \textbf{256         } & 256            & 120       & 120         & 616 \\

Through. (GOPS)    & \textbf{4.2        } & 3.1         & 1.2       & 0.4         & \textbf{34.6        } & 22.2$^\dagger$ & 1.7       & 1.8         & 4.6 \\

Speedup            & \textbf{3.5$\times$} & 2.6$\times$ & 1$\times$ & 0.3$\times$ & \textbf{20.7$\times$} & 13.3$\times$   & 1$\times$ & 1.1$\times$ & 2.8$\times$ \\

Power (W)          & \textbf{0.11       } & 0.07        & 55        & 98          & \textbf{1.1         } & 16             & 61        & 65          & 155 \\

EDP (pJ$\times$ns) & \textbf{6.0        } & 7.1         & 38k       & 1M          & \textbf{1.0         } & 57.4           & 36k       & 27k         & 9k\\

\bottomrule
\end{tabular}%
}
\end{table*}

\subsection{State-of-the-art comparison} \label{sec:sota_comparison}
\re{The proposed processor is benchmarked against two state-of-the-art ASIPs and optimized CPU and GPU baselines.

\begin{itemize}[itemsep=0em, leftmargin=*]
    \item \re{\textbf{CPU}: A state-of-the-art multi-threaded implementation of DAGs} \cite{shah2021graphopt} is benchmarked on an Intel(R) Xeon Gold 6154 CPU with 18 cores using the open-sourced code\footnote{https://github.com/nimish15shah/GRAPHOPT}.
    \item \textbf{GPU}: The SpTRSV implementation \cite{naumov2011parallel} from the cuSPARSE library is benchmarked on an RTX 2080Ti GPU. In the absence of a GPU library for PC, we implemented a layer-wise parallelization in CUDA \cite{cuda} that is inspired by the cuSPARSE SpTRSV implementation.
\end{itemize}}
\re{\subsubsection{Comparison using PC and SpTRSV}\label{sec:dpu_comparison}
For the PC and SpTRSV workloads in tables~\ref{tab:workload_details}(a) and (b), the performance of the obtained min-EDP configuration of our processor is compared with a SoTA processor targeting irregular DAGs called the DAG processing unit (DPU) \cite{shah2021dpu, shah20219}, in addition to CPU and GPU.

\textbf{DPU}: 
DPU, the prior architecture generation of this line of research, has an architectural approach similar to fig.~\ref{fig:async_vs_sync}(a).
A configuration of DPU with a similar area, number of registers, arithmetic format, and on-chip memory bandwidth (112.5GB/s) as the current work is synthesized in a 28nm technology from the open-source RTL\footnote{https://github.com/nimish15shah/DPU\_DAG\_Processing\_Unit}. 

\textbf{Results}: Fig.~\ref{fig:throughput_all_w}(a) shows the throughput of individual workloads and table \ref{tab:sota_comparison} summarizes the results. 
\name{} outperforms DPU for all workloads except \texttt{bnetflix} and \texttt{sieber}, with an average speedup of 1.4$\times$, 4.2$\times$, and 10.5$\times$ over DPU, CPU, and GPU, respectively, while achieving 15\% better EDP than DPU.

\textbf{Analysis}: Although operating at higher frequencies, the CPU and GPU underperform specialized architectures due to severe inefficiency in the cache hierarchies due to the irregular fine-grained accesses (as discussed in \S \ref{sec:introduction}). Furthermore, due to the relatively small size of the workloads, there is not enough parallel work to amortize the synchronization and communication overheads of multiple cores. DPU addresses these problems leading to relatively modest gains over it. The major reasons for the speedup of \name{} over DPU are as follows. 
\begin{itemize} [itemsep=0em, leftmargin=*]
    \item DPU does not enable the reuse of intermediate variables within the datapath, which is achieved in this work through PE trees. Consequently, this work can utilize more PEs than DPU for the same number of input/output ports of the datapath and leads to fewer register file accesses.
    \item In DPU, 43\% of the load requests result in bank conflicts, severely underutilizing the on-chip data memory bandwidth. To reduce the impact on throughput, aggressive hardware prefetching is performed to overlap memory requests with computations, at the cost of higher hardware complexity. In contrast, this work simplifies the hardware by 
    considerably eliminating the bank conflicts altogether by limiting the irregular accesses to register banks and through advanced bank mapping during compilation. As discussed in \S \ref{sec:async_vs_sync}, such a compile-time conflict avoidance is difficult for DPU due to asynchronous execution of cores leading to an unpredictable timing of load requests.
\end{itemize}}
\re{\subsubsection{Comparison using large PCs} \label{sec:spu_comparison}
To demonstrate the scalability of this work, a large configuration of the proposed processor is benchmarked with large PCs with up to 3.3M nodes (table \ref{tab:workload_details}(c)) and compared with the sparse processing unit (SPU) \cite{dadu2019towards}, the CPU baseline from the SPU work (CPU$_{SPU}$), CPU, and GPU.
\begin{itemize} [itemsep=0em, leftmargin=*]
    \item \textbf{SPU}: SPU is a coarse-grained reconfigurable array (CGRA)-like architecture tuned for sparse workloads. Since the code is not open-sourced, we estimate the throughput based on the speedups reported over its CPU baseline.
    \item \textbf{CPU$_{SPU}$}: For estimating the throughput of SPU and for a fair comparison of SPU speedup with our work, the CPU baseline from SPU is benchmarked\footnote{The CPU$_{SPU}$ baseline code is provided by the SPU authors}.
    \item \textbf{\name{} (L)}: For the large workloads, our processor is synthesized with a larger on-chip data memory (2MB) and 256 registers per bank. Unlike previous experiments, the instructions do not fit in the on-chip memory anymore, and are streamed from external interface with a bandwidth of 64 GB/s. Since SPU assumes a 256 GB/s memory interface and performs batch execution, our system is benchmarked with 4 cores (with 64GB/s requirement per core) for a fair comparison. The parallel cores can either perform batch execution (used for benchmarking) or execute different DAGs. Like the SPU experiments, the power consumption of an external memory and memory controller are not included in the measurements.
\end{itemize}

\textbf{Results}: The per-workload throughput is shown in fig.~\ref{fig:throughput_all_w}(b) and the overall results are summarized in table \ref{tab:sota_comparison}. With a similar area and external bandwidth consumption, the proposed processor outperforms prior work, with an overall speedup of 1.6$\times$, 20.7$\times$, 19.2$\times$, and 7.5 $\times$ over SPU, CPU$_{SPU}$, CPU, and GPU, respectively. Furthermore, the speedup is achieved at significantly lower power (1.1W) compared to SPU (16W), greatly improving EDP.



\textbf{Analysis}: The large DAGs are able to better utilize GPU, leading to higher throughput than CPU, but also consume higher power. The CPU and GPU still underperform compared to specialized hardware due to the reasons discussed in \S \ref{sec:introduction}. \name{} (L) consumes significantly lower power than SPU due to the following main architectural differences:
\begin{itemize} [itemsep=-0em, leftmargin=*]
    \item SPU does not have a register file leading to significantly higher memory accesses compared to this work.
    \item Although, in general, SPU supports the reuse within the CGRA-like datapath, it could not be fully utilized for unstructured DAGs due to the lack of a frequently recurring pattern for which the CGRA connectivity can be fixed. 
    \item SPU aggressively reorders memory requests in hardware to properly utilize the on-chip memory bandwidth in the presence of irregular requests, at the expense of higher hardware complexity. On the other hand, this work considerably eliminates the bank conflicts altogether by limiting the irregular accesses to register banks and through advanced bank mapping during compilation.
\end{itemize}
}


This way, the hardware-software co-optimization presented for \name{} results in efficient parallel execution of irregular DAGs for emerging embedded applications.



\section{Additional related works} \label{sec:related_works}
\re{\textbf{Specialized dataflow}:}
MAERI \cite{kwon2018maeri}, an architecture with flexible dataflow for neural networks, is the most similar to this work, as it also uses a tree-based datapath of PEs. 
But, the controller and the interconnection networks are designed specifically for neural networks, and do 
not natively support arbitrary DAGs. Due to the regularity in NNs, the mapping of operations and routing of operands get significantly simplified compared to arbitrary DAGs. The regularity also limits bank conflicts, which get exacerbated due to irregularity. 

\re{Clark et al.~proposed approaches related to this work, in which frequent recurring subgraphs are identified in a DAG \cite{clark2005automated}, \cite{clark2004application} to determine the most promising datapath structure.
However, the number of ports of the datapath (4 inputs, 2 outputs) are severely limited due to the use of a monolithic register file, limiting hardware parallelism. This work addresses that by using a banked register file and appropriate interconnections, allowing a highly parallel datapath with up to 64 read/write ports.}

\re{Chainsaw \cite{sharifian2016chainsaw} decomposes DAGs into \textit{chains} instead of trees, which is also a promising alternative as multi-input nodes can also be unfolded as chains. 
However, the inter-chain communication, which is frequent for irregular DAGs, happens via a central register file accessed with a bus. This would become a major bottleneck as fig.~\ref{fig:interconnect_dse} shows that a multi-ported register file and high-bandwidth interconnection used in this work are critical for good performance.

SEED\cite{nowatzki2015exploring} combines a dataflow architecture with a general-purpose processor. It focuses on small DAGs within nested loops, which typically contain a high proportion of conditional nodes, while our work focuses on large DAGs,
requiring a different approach based on high-bandwidth interconnects and a banked register file.}

\re{\textbf{Operand networks:} The input/output interconnects of our work can be classified as \textit{scalar operand networks} as discussed in \cite{taylor2003scalar}, however we closely couple operand networks with a spatial tree-based datapath, which significantly complicates the spatial and temporal mapping. 
The conflict-avoiding bank mapping given the networking constraints is also novel.}

\re{
\vspace{5pt}
\noindent \textbf{FPGA:} The work in \cite{8945724} targets PCs through fully-spatial mapping on FPGA. However, since large PCs cannot be completely spatially mapped, their approach is limited to only small DAGs with fewer than 500 nodes.}

\vspace{5pt}
\noindent \textbf{CGRA:} Coarse-grained reconfigurable array (CGRA) architectures like HRL \cite{7446059}, Plasticine \cite{8192487}, DySER \cite{6235947}, \re{CGRA Express \cite{park2009cgra}} can execute DAGs with a spatial datapath. However, existing CGRAs suffer from several problems. 
\begin{itemize}[itemsep=0em , leftmargin=*]
    \item Most CGRAs either have a local register file within a PE or do not have a register file at all. Eg., DySER does not have a register file in the PEs and the inputs/outputs are sequentially accessed from a global register file, which would severely limit the throughput. 
    \re{CGRA Express \cite{park2009cgra} uses a local register file per PE and also has register bypass networks to dynamically fuse operations with the same goal as the PE trees in this work. An advantage of using PE trees is that $X$ PEs can operate with $X+1$ register read ports, while in CGRA Express, $X$ PEs would require $2X$ ports. Furthermore, experiments in \S \ref{sec:interconnect_dse} shows that a one-to-one connection of PE and banks would lead to frequent conflicts, revealing that local register files (similar to a one-to-one connection) would be insufficient and more complex connectivity between the PEs and register banks is essential for good performance.}
    \item The spatial connectivity of CGRAs is not reconfigured every cycle typically, but is fixed for the entire execution. Given that a large DAG cannot be entirely mapped spatially, it has to be decomposed into one recurring pattern, which is difficult to find due to irregularity.
    \item CGRAs normally use a 2D mesh. But, as described in \S \ref{sec:systolic_vs_tree}, unstructured DAGs cannot fully utilize a 2D array.
\end{itemize}



\vspace{5pt}
\noindent \textbf{Graph analytics:} In recent years, several accelerators like Fifer \cite{nguyen2021fifer}, PolyGraph \cite{dadu2021polygraph}, Graphicionado \cite{ham2016graphicionado}, \cite{gui2019survey}, with hardware support for irregularity have been proposed for graph kernels like traversal, node ranking, clustering, etc.
In graph analytics, the ordering of nodes/edges cannot be determined at compile time, limiting the scope of compile-time optimizations. On the other hand, the node ordering in our target computation DAGs is fixed based on the dependencies, which are known at compile time. This fact is utilized extensively in this work to perform mapping of nodes to PEs, conflict-aware bank allocation, pipeline-aware reordering, etc. 
This way, the hardware is simplified by assigning several responsibilities to the compiler, \rtwo{but excludes general graph analytic operations from the scope of this work}. 

\vspace{5pt}
\noindent \textbf{Spare linear algebra:} A growing number of hardware solutions are being designed for sparse linear algebra, like Sparse-TPU \cite{he2020sparse}, SpArch \cite{zhang2020sparch}, SparseP \cite{giannoula2022sparsep}, etc. Some specifically target sparsity in deep learning algebra, e.g, SNAP \cite{9310233}, Sticker \cite{8890710}, \cite{dave2021hardware}. These works mainly focus on sparse matrix-matrix or matrix-vector multiplications (SpMM and SpMV). But SpTRSV has a different kind of parallelism compared to SpMM and SpMV due to its \textit{inductive} nature, which is discussed in detail in REVEL \cite{weng2020hybrid}. Put simply, in SpMV and SpMM, outputs only depend on inputs, while in SpTRSV, outputs can depend on the previous outputs, creating possibly long chains of producer-consumer dependencies. REVEL \cite{weng2020hybrid} provides a solution for dense matrices.
Sparsity exacerbates the parallelization difficulties by introducing irregularity, which is addressed in this work. \rtwo{There still remains scope for a more general solution that does not utilize a DAG-specific (in other words, sparsity-pattern specific) compilation developed in this work, allowing the sparsity pattern to change with every execution.} 















\vspace{5pt}
\noindent \textbf{Compilers:} 
\re{The greedy block generation in step 1 (\S \ref{sec:block_decomposition}) is highly related to the work in \cite{clark2006scalable}, and their advanced algorithm that can achieve optimal mapping for a general spatial datapath. However, their experiments reported a modest speedup of 10\% over greedy heuristics, suggesting a limited opportunity for further improvement over this work.}
CGRA mappers like \cite{canesche2020traversal}, 
\cite{hamzeh2013regimap}, \cite{walker2019generic}, \cite{chen2014graph}, \cite{nowatzki2018hybrid}, \cite{nowatzki2014scheduling}, can map a DAG to generic spatial datapaths, including trees. However, they can handle relatively small DAGs with up to 500 nodes, which are supposed to be fully mapped to the datapath. In contrast, this work can handle more than 100k nodes by limiting the focus to trees.
Moreover, these works usually considered a local/no register file, making them unsuitable for a global, banked file.

Like this work, \cite{cong2016scalable} also developed a DAG mapping algorithm for an architecture with a tree of accelerator blocks (instead of PEs) \cite{cong2012charm}, but the similarity ends there. Since the target
was a larger hardware granularity, i.e, accelerator blocks instead of PEs, they did not have to consider the constraints of a banked register file connected with a custom interconnection topology, like this work.



VLIW compilers that consider partitioned register files like \cite{ellis1986bulldog}, \cite{jang1998code}, and \cite{ozer1998unified}, share several similarities with the compiler of this work. Specifically, the bottom-up greedy (BUG) algorithm \cite{ellis1986bulldog} can compile for independent register banks and arbitrary interconnection between the banks and PEs.
Yet, 
it enforces a limitation that the output of a PE cannot be connected to another PE but can only connect to a register bank.
This limits the design space to only an array of PEs
(i.e. our template with $D$=1) and excludes the PE trees used in this work. As noted in \cite{ellis1986bulldog}, the algorithm cannot be easily modified to ease this constraint. Our compiler, on the other hand, targets PE trees but limits the interconnection design space to have at least one crossbar.

\section{Conclusion} \label{sec:conclusion}
This paper presents \name{}, a specialized processor template designed for energy-efficient parallel execution of irregular DAGs. The architecture is equipped with a tree-based datapath of processing elements enabling immediate data reuse. Instead of a multi-ported global register file, parallel register banks with independent addressing provide the required bandwidth. The instruction overhead of independent register addresses is eased with an automatic writing scheme. The optimized interconnect networks between the datapath and the register banks enable flexible routing of data required for irregular data accesses. The architecture is made programmable with a variable-length VLIW-like instruction set.

A targeted compiler is developed to map DAGs to \name{}, which maximizes the datapath utilization, while minimizing stalls due to bank conflicts and pipeline hazards. This way, a cohesive hardware-software co-optimization approach is developed to address the complexities arising from irregularity and simplify the hardware for energy efficiency.

Finally, a design space exploration is performed to identify the configuration with minimal EDP. The optimal design achieved speedup over prior works while consuming lower EDP.
Thus, this paper demonstrates the effectiveness of hardware-software co-optimization for irregular DAG execution, enabling emerging energy-constrained applications.


\section*{Acknowledgment}
This work has been supported by the European Union's European Research Council (ERC) project Re-SENSE under grant agreement ERC-2016-STG-715037, and received funding from the Flemish Government (AI Research Program) and KU Leuven. We would like to thank Laura I. Galindez Olascoaga (UC Berkley) for the insightful discussions of probabilistic circuits and benchmarks; Vidushi Dadu and Tony Nowatzki (UC Los Angeles) for providing CPU$_{SPU}$ baseline implementation; and Yitao Liang and Guy Van den Broeck (UC Los Angeles) for providing probabilistic circuit workloads.

\clearpage
\input{artifact_appendix}

\clearpage

\bibliographystyle{IEEEtranS}
\bibliography{ref}

\end{document}

%% file: algo_step1.tex
\newcommand\mycommfont[1]{\scriptsize\ttfamily\color{red}{#1}}
\SetCommentSty{mycommfont}

\SetKwComment{tcp}{$\triangleright$ }{}%
\SetKwComment{tcc}{$\triangleright$ }{}%



\SetAlFnt{\scriptsize}
\SetAlCapFnt{\scriptsize}
\SetAlCapNameFnt{\scriptsize}

\begin{algorithm}
\let\oldnl\nl
\newcommand{\nonl}{\renewcommand{\nl}{\let\nl\oldnl}}

\DontPrintSemicolon 
\tcc{Decompose into blocks (section \S \ref{sec:block_decomposition})}
\KwIn{
    $G(V,E)$: Binarized DAG where $V$ is the set of nodes and $E$ is the set of edges. \depth{}, \num{}: The depth and number of trees in the datapath.
}
\KwOut{$\mi{\mathcal{B}}$: Set of blocks\newline
    }

$\mi{\mathcal{B}} \gets \varnothing$, $\mi{\mathcal{D}_{sch}} \gets \varnothing$, $\mi{dfs\_order} \gets$ dfs($G$), $\mi{done\_nodes} \gets \varnothing$


$\mi{combos} \gets$ possible\_depth\_combinations($D$, $T$) \tcp{See fig.~\ref{fig:compiler_step_1}(d)}



$\mathit{curr\_source\_nodes} \gets$ Nodes with no incoming edges

\nonl \;

\While{$\mi{len(done\_nodes)} \mathrm{\; \neq \;} \mi{len(V)}$}
{
    \For(\tcp*[h]{Find sched. subgraphs}){$n \mathrm{\; in \;} \mi{curr\_source\_nodes}$}
    {
        $\mi{temp\_\mathcal{D}_{sch}} \gets$ find\_schedulable\_subg($n$)
    
        $\mi{\mathcal{D}_{sch}} \gets \mi{\mathcal{D}_{sch}} \cup \mi{temp\_\mathcal{D}_{sch}}$
    }

    \nonl \;
    
    \tcp{Choose the \textit{best} subgraphs to form a block}
    $\mi{best\_fitness} \gets$ 0, $\mi{block} \gets \varnothing$

    \For{$\mi{combo} \mathrm{\; in \;} \mi{combos}$} 
    {
        $\mi{temp\_block} \gets \varnothing$ 

        \tcc{$\mi{combo}$ is a list of subgraph depths that can be combined in a block. For example, $\mi{combo}$=[2, 1, 1] represents the third combination in fig.~\ref{fig:compiler_step_1}(d).}
        $d_{max} \gets \mi{combo}[0]$
        
        $\mi{largest\_subg} \gets$
        get\_largest\_subg($\mi{\mathcal{D}_{sch}}$, $d_{max}$)
        
        $\mi{temp\_block.add(largest\_subg}$)

        \For{$d \mathrm{\; in \;} \mi{combo}[1 : ]$} 
        {
            \tcc{Iterate over subgraphs in $\mi{\mathcal{D}_{sch}}$ to find a suitable subgraph of depth $d$}
            
            $\mi{subg\_of\_depth\_d} \gets$ \newline get\_fittest\_subg($\mi{\mathcal{D}_{sch}}$, $d$, $\mi{largest\_subg}$, $\mi{dfs\_order}$)
            
            $\mi{temp\_block.add(subg\_of\_depth\_d}$)
        }
        
        $\mi{temp\_fitness} \gets$ get\_fitness($\mi{temp\_block}$)
        
        \If{$\mi{temp\_fitness} > \mi{best\_fitness}$}
        {
             $\mi{block} \gets \mi{temp\_block}$, $\mi{best\_fitness} \gets \mi{temp\_fitness}$
        }
        
    }
    
    \nonl \;
    
    \tcc{Remove all the subgraphs from $\mi{\mathcal{D}_{sch}}$ that contain a node allocated to $\mi{block}$.}
    \For{$\mi{subg}  \mathrm{\; in \;} \mi{\mathcal{D}_{sch}}$}
    {
        \If{$\mathrm{any \; node \; from \;} \mi{block} \mathrm{\;is \; in \;} \mi{subg}$}
        {
            $\mi{\mathcal{D}_{sch}.remove(subg}$)
        }
    }
    
    \nonl \;
    
    $\mi{\mathcal{B}.add(block}$), $\mathit{curr\_source\_nodes} \gets$ Nodes of $\mi{block}$

    $\mi{done\_nodes}.add$(Nodes of $\mi{block}$)
}
\Return{$\mi{\mathcal{B}}$}

\caption{\rtwo{Step 1  $\langle G(V, E), D, T
\rangle$
    }
    }
\label{alg:step_1}

\end{algorithm}

%% file: algo_step2.tex

\begin{algorithm*}

\SetAlFnt{\scriptsize}
\SetAlCapFnt{\scriptsize}
\SetAlCapNameFnt{\scriptsize}

\let\oldnl\nl
\newcommand{\nonl}{\renewcommand{\nl}{\let\nl\oldnl}}

\DontPrintSemicolon 
\tcc{PE and register bank mapping(section \S \ref{sec:pe_bank_allocation})}
\KwIn{
    $G(V,E)$: Binarized DAG where $V$ is the set of nodes and $E$ is the set of edges. \depth{}: The depth of PE trees in the datapath. \bank{}: The number of register banks. $\mi{output\_connectivity}$: Information of which PE's output is connected to which banks (Note that the input connectivity is assumed to be a crossbar). $\mi{\mathcal{B}}$: The set of blocks generated in step 1.

}
\KwOut{$\mi{\mathcal{M}_{p}}$: Mapping of nodes to PEs. $\mi{\mathcal{M}_{b}}$: Mapping of input/output nodes to banks.
    }
    
$\mi{\mathcal{M}_{p}} \gets \{\}$, $\mi{\mathcal{M}_{b}} \gets \{\}$, $\mi{\mathcal{S}_{p}} \gets \{\}$, $\mi{\mathcal{S}_{b}} \gets \{\}$

$\mi{io\_nodes} \gets$ Nodes that serve as inputs/outputs of blocks

\For(\tcp*[h]{Initialize the sets of compatible PEs and banks}){$\mi{block} \mathrm{\; in \;} \mi{\mathcal{B}}$, $\mi{subg} \mathrm{\; in \;}  \mi{block}$, $\mi{node} \mathrm{\; in \;} \mi{subg}$}
{
            \tcc{Determine which PEs are compatible based on the position of the node in the subgraph. For example, only the top PE is compatible for mapping node $a$ in fig.~\ref{fig:bank_allocation}.}
            $d \gets$ get\_max\_depth\_from\_sources($\mi{subg}$, $\mi{node}$)
            
            $\mi{compatible\_PEs} \gets$ PEs at depth $d$ in the datapath
            
            $\mi{\mathcal{S}_{p}}[\mi{node}] \gets \mi{compatible\_PEs}$
        
            \If{$\mi{node} \mathrm{\; in \;} \mi{io\_nodes}$}
            {
                $\mi{\mathcal{S}_{b}}[\mi{node}] \gets \mi{output\_connectivity}[\mi{compatible\_PEs}]$ \tcp{Banks that are writable from the compatible PEs}
            }
}

\tcc{A dictionary to find the node with the minimum number of compatible banks in $O$(\bank{}) time irrespective of the DAG size. Key: The number of compatible banks, Value: The set of nodes.}
$\mi{\mathcal{M}_{nodes}} \gets$ \{$b$:$\varnothing$\ \texttt{for} $b$ in [0,1,..,\bank{}]\}

\For{$\mi{node} \mathrm{\; in \;} \mi{io\_nodes}$}
{
    $\mi{num\_banks} \gets $ $len(\mi{\mathcal{S}_{b}}[\mi{node}])$
    
    $\mi{\mathcal{M}_{nodes}[\mi{num\_banks}]}.add(\mi{node})$
}

\tcp{Iteratively map nodes from $\mi{io\_nodes}$ to PEs and banks}

$\mi{done\_nodes} \gets \varnothing$

\While{$\mi{len(done\_nodes)} \mathrm{\; \neq \;} \mi{len(io\_nodes)}$}
{   
    \tcc{Choose a node with the minimum number of compatible banks. If there are multiple, choose any one randomly.}
    \For{$b \mathrm{\; in \;} [0,...,B]$}
    {
        \If{$\mi{\mathcal{M}_{nodes}[b]} \mathrm{\; \neq \;} \varnothing$}
        {
            $\mi{chosen\_node} \gets \mi{\mathcal{M}_{nodes}[b]}.pop(\mi{random})$
            
            \texttt{break}
        }
    }
    
    $\mi{simul\_wr} \gets$ Nodes written simultaneously with $\mi{chosen\_node}$
    
    $\mi{simul\_rd} \gets$ Nodes read simultaneously with $\mi{chosen\_node}$
    
    
    \tcp{Bank mapping}
    \If  (\tcp*[h]{A compatible bank exists}) {$\mi{\mathcal{S}_{b}}[\mi{chosen\_node}] \mathrm{\; \neq \;} \varnothing$}
    {
        $\mi{chosen\_bank} \gets \mi{\mathcal{S}_{b}}[\mi{chosen\_node}].pop(\mi{random})$ \tcp{In practice, choosing randomly achieves Objective J}
    }
    \Else (\tcp*[h]{No compatible bank left})
    {
        \tcc{Bank conflicts will happen. Minimize the number of conflicts by choosing a bank that is least contented.}
        $\mi{chosen\_bank} \gets$ A bank that is allocated to the least number of nodes in $\mi{simul\_wr}$ and $\mi{simul\_rd}$ sets
    }
    
    \tcp{PE mapping}
    $\mi{desirable\_PEs} \gets$ (PEs that can write to $\mi{chosen\_bank}$) $\cap$ $\mi{\mathcal{S}_{p}}[\mi{chosen\_node}]$
    
    \If(\tcp*[h]{No bank conflict while writing.}){$\mi{desirable\_PEs} \mathrm{\; \neq \;} \varnothing$}
    {
        $\mi{chosen\_PE} \gets \mi{desirable\_PEs}.pop(\mi{random})$
    }
    \Else (\tcp*[h]{A bank conflict while writing, but the $\mi{chosen\_PE}$ is according to constraint E.})
    {
        $\mi{chosen\_bank} \gets \mi{\mathcal{S}_{p}}[\mi{chosen\_node}].pop(\mi{random})$
    }
    
    \tcp{Intra-block compatibility update}
    
    Remove $\mi{chosen\_bank}$ from $\mi{\mathcal{S}_{b}}$ of $\mi{simul\_wr}$ nodes, which by definition are from the same block as $\mi{chosen\_node}$, and update $\mi{\mathcal{M}_{nodes}}$ accordingly.
    
    \For{$node \mathrm{\; in \; the \; block \; of \;} \mi{chosen\_node}$}
    {
        Update $\mi{\mathcal{S}_{p}}[\mi{node}]$ according to constraint $E$
        
        (If $\mi{node}$ in $\mi{io\_nodes}$) Update $\mi{\mathcal{S}_{b}}[\mi{node}]$ according to the updated $\mi{\mathcal{S}_{p}}[\mi{node}]$ based on the $\mi{output\_connectivity}$. Also, update $\mi{\mathcal{M}_{nodes}}$ accordingly.
    }
    \tcp{Inter-block compatibility update}
    Remove $\mi{chosen\_bank}$ from $\mi{\mathcal{S}_{b}}$ of $\mi{simul\_rd}$ nodes, and update $\mi{\mathcal{M}_{nodes}}$ accordingly.
    
    \nonl \;
    
    $\mi{\mathcal{M}_{p}[chosen\_node]} \gets \mi{chosen\_PE}$,     $\mi{\mathcal{M}_{b}[chosen\_node]} \gets \mi{chosen\_bank}$, $\mi{\mathcal{S}_{b}.remove[chosen\_node]}$, $\mi{\mathcal{S}_{p}.remove[chosen\_node]}$
    
    $\mi{done\_nodes.add(chosen\_node)}$
}

\nonl \;

\tcp{Allocate PEs to the rest of the nodes (i.e. non $\mi{io\_nodes}$) just as above but without bank-mapping related steps.}

\Return{$\mi{\mathcal{M}_{p}, \mi{\mathcal{M}_{b}}}$}

\caption{\rtwo{Step 2  $\langle G(V, E), D, B, \mi{output\_connectivity}, \mi{\mathcal{B}}
\rangle$
    }
    }
\label{alg:step_2}

\end{algorithm*}  



%% file: artifact_appendix.tex
\appendix
\section{Artifact Appendix}

\subsection{Abstract}
To reproduce the performance results of our processor, the following items are open-sourced:
\begin{itemize}
    \item SystemVerilog-based microarchitectural register transfer level (RTL) model of \name{} and a testbench.
    \item The custom compiler implemented in Python
    \item The input DAGs used in the experiments
\end{itemize}

\textbf{Workflow summary}:
The python-based compiler takes the DAG benchmarks as inputs and generates binary programs for each benchmark. The SystemVerilog testbench reads the binary programs and the data input files, and executes the RTL simulation (with the proprietary Synopsys VCS simulator) to evaluate the processor's throughput. Finally, charts are plotted according to the obtained results. All these steps are automated with a python script.

\textbf{Expected results}:
The experiments report the breakdown of the different types of instruction (fig. \ref{fig:instr_stat_all_w}) and the throughput achieved by the processor (fig. \ref{fig:throughput_all_w}(a)).

\subsection{Artifact check-list (meta-information)}

{\small
\begin{itemize}
  \item {\bf Algorithm: }A compilation algorithm is presented to map a DAG to the proposed processor.
  \item {\bf Compilation: }The custom compiler is included with the codebase.
  \item {\bf Data set: }The target DAG workloads are included with the codebase.
  \item {\bf Run-time environment: }The code is tested on two Linux distributions (CentOS and Ubuntu).
  \item {\bf Hardware: }The processor microarchitecture is modeled at register transfer level (RTL) with SystemVerilog. Ths Synopsys VCS simulator is used to perform the RTL simulation.
  
  \item {\bf Execution: }The execution runs for approximately 3-4 hours.
  
  \item {\bf Metrics: }The latency reported by the cycle accurate RTL simulation is used to compute the throughput of the processor.
  
  \item {\bf Output: }The outputs of the experiments are: \begin{enumerate}
      \item A plot showing the breakdown of the instruction categories as reported in fig.~\ref{fig:instr_stat_all_w}.
      \item A plot showing the throughput of the processor as reported in fig.~\ref{fig:throughput_all_w}(a). The plot also shows throughput of other platforms for comparison, but they are NOT evaluated in this codebase.
  \end{enumerate}
  Note that the large DAGs ("Large PC") from table \ref{tab:workload_details} are not evaluated due to long runtime ($>$24hours).
  \item {\bf Experiments: } The experiments are run through a bash script.
  
  \item {\bf How much disk space required (approximately)?: }Less than 200MB.
  
  \item {\bf How much time is needed to prepare workflow (approximately)?: }10 minutes, if the software dependencies are already installed.
  
  \item {\bf How much time is needed to complete experiments (approximately)?: }7 hours.
  
  \item {\bf Publicly available?: }Yes, at \url{https://github.com/nimish15shah/DAG\_Processor}.
  
  \item {\bf Code licenses (if publicly available)?: }MIT license.
  
  \item {\bf Data licenses (if publicly available)?: }The input workloads (provided with the codebase for quick experimentation) are publicly available in the UCLA StarAI circuit model zoo (\url{https://github.com/UCLA-StarAI/Circuit-Model-Zoo/tree/master/psdds}) and the SuiteSparse matrix collection (\url{https://sparse.tamu.edu/}).
  
  
  \item {\bf Archived (provide DOI)?: }Yes, available on Zenodo. \\ \url{https://doi.org/10.5281/zenodo.7020493}
\end{itemize}
}

\subsection{Description}

\subsubsection{How to access}
The artifact is made available at \url{https://github.com/nimish15shah/DAG\_Processor} and archived on Zenodo at \url{https://doi.org/10.5281/zenodo.7020493}. The disk space requirement is less than 200MB.

\subsubsection{Hardware dependencies}
No special hardware is required other than a general purpose CPU running a Linux OS.

\subsubsection{Software dependencies}
The following softwares are required:
\begin{itemize}
    \item The proprietary Synopsys VCS simulator for SystemVerilog simulation. More information at \url{https://www.synopsys.com/verification/simulation/vcs.html}.
    \item bash version 4 (or zsh).
    \item (Optional) Anaconda. Please find the Linux installation guide at \url{https://docs.anaconda.com/anaconda/install/linux/\#installation}.
\end{itemize}
The experiments are tested on Linux distributions (CentOS and Ubuntu).

\subsubsection{Data sets}
The input workloads are provided with the codebase for quick experimentation. They are publicly available in the UCLA StarAI circuit model zoo (\url{https://github.com/UCLA-StarAI/Circuit-Model-Zoo/tree/master/psdds}) and the SuiteSparse matrix collection (\url{https://sparse.tamu.edu/}).

\subsection{Installation}
Please run the following commands for setting up the project.\\

\noindent \textbf{With Anaconda: }

\begin{minted}
[ fontsize= \scriptsize,
 breaklines, showspaces, space=~
]
{bash}
git clone git@github.com:nimish15shah/DAG_Processor.git
cd DAG_Processor
conda create --name DAGprocessor --file conda-linux-64.yml
\end{minted}



\noindent \textbf{Without Anaconda: }

\begin{minted}
[ fontsize= \scriptsize,
 breaklines, showspaces, space=~
]
{bash}

git clone git@github.com:nimish15shah/DAG_Processor.git
cd DAG_Processor
python3 -m venv venv_DAGprocessor
./venv_DAGprocessor/bin/activate
pip install --upgrade pip
pip install -r requirements.txt

\end{minted}


\subsection{Experiment workflow}
The experiments are launched with the following script.\\

\noindent \textbf{With Anaconda (recommended): }
\begin{minted}
[ fontsize= \scriptsize,
 breaklines, showspaces, space=~
]
{bash}
conda activate DAGprocessor
./run.sh 
\end{minted}
\noindent \textbf{Without Anaconda: }

\begin{minted}
[ fontsize= \scriptsize,
 breaklines, showspaces, space=~
]
{bash}
./run.sh noconda
\end{minted}

The script performs the following steps:
\begin{itemize}
    \item The python-based compiler takes the DAG benchmarks as input and generates binary programs for each benchmark.
    \item The SystemVerilog testbench reads the binary programs and the data input files generated by the compiler, and executes the RTL simulation (with the proprietary Synopsys VCS simulator) and generates the latency log.
    \item Finally the charts are plotted according to the data collected during the compilation and RTL simulations.
\end{itemize}
\subsection{Evaluation and expected results}
The expected results of the experiments are: \begin{enumerate}
      \item A plot at \\ \texttt{./out/plots/instruction\_breakdown.pdf} showing the breakdown of the instruction categories as reported in fig.~\ref{fig:instr_stat_all_w}.
      \item A plot at \texttt{./out/plots/throughput.pdf} showing the throughput of the processor as reported in fig.~\ref{fig:throughput_all_w}(a). The plot also shows throughput of other platforms for comparison, but they are NOT evaluated in this artifact.
  \end{enumerate}
  
  Note 1: The large DAGs ("Large PC") from table \ref{tab:workload_details} are not evaluated due to large experimental runtime ($>$24hours).
  
  Note 2: There could be slight variations in the results in every run due to some randomizations in the compiler.
